\documentstyle[twoside,epsf]{article}

\catcode`\@=11
\long\def\@makefntext#1{
\protect\noindent \hbox to 3.2pt {\hskip-.9pt  
$^{{\eightrm\@thefnmark}}$\hfil}#1\hfill}		

\def\@makefnmark{\hbox to 0pt{$^{\@thefnmark}$\hss}}	
	
\def\ps@myheadings{\let\@mkboth\@gobbletwo
\def\@oddhead{\hbox{}
\rightmark\hfil\eightrm\thepage}   
\def\@oddfoot{}\def\@evenhead{\eightrm\thepage\hfil
\leftmark\hbox{}}\def\@evenfoot{}
\def\sectionmark##1{}\def\subsectionmark##1{}}

\oddsidemargin=\evensidemargin
\newcounter{sectionc}\newcounter{subsectionc}\newcounter{subsubsectionc}
\renewcommand{\section}[1] {\vspace{12pt}\addtocounter{sectionc}{1} 
\setcounter{subsectionc}{0}\setcounter{subsubsectionc}{0}\noindent 
	{\tenbf\thesectionc. #1}\par\vspace{5pt}}
\renewcommand{\subsection}[1] {\vspace{12pt}\addtocounter{subsectionc}{1} 
	\setcounter{subsubsectionc}{0}\noindent 
	{\bf\thesectionc.\thesubsectionc. {\kern1pt \bfit #1}}\par\vspace{5pt}}
\renewcommand{\subsubsection}[1] {\vspace{12pt}\addtocounter{subsubsectionc}{1}
	\noindent{\tenrm\thesectionc.\thesubsectionc.\thesubsubsectionc.
	{\kern1pt \tenit #1}}\par\vspace{5pt}}
\newcommand{\nonumsection}[1] {\vspace{12pt}\noindent{\tenbf #1}
	\par\vspace{5pt}}

\newcounter{appendixc}
\newcounter{subappendixc}[appendixc]
\newcounter{subsubappendixc}[subappendixc]
\renewcommand{\thesubappendixc}{\Alph{appendixc}.\arabic{subappendixc}}
\renewcommand{\thesubsubappendixc}
	{\Alph{appendixc}.\arabic{subappendixc}.\arabic{subsubappendixc}}

\renewcommand{\appendix}[1] {\vspace{12pt}
        \refstepcounter{appendixc}
        \setcounter{figure}{0}
        \setcounter{table}{0}
        \setcounter{lemma}{0}
        \setcounter{theorem}{0}
        \setcounter{corollary}{0}
        \setcounter{definition}{0}
        \setcounter{equation}{0}
        \renewcommand{\thefigure}{\Alph{appendixc}.\arabic{figure}}
        \renewcommand{\thetable}{\Alph{appendixc}.\arabic{table}}
        \renewcommand{\theappendixc}{\Alph{appendixc}}
        \renewcommand{\thelemma}{\Alph{appendixc}.\arabic{lemma}}
        \renewcommand{\thetheorem}{\Alph{appendixc}.\arabic{theorem}}
        \renewcommand{\thedefinition}{\Alph{appendixc}.\arabic{definition}}
        \renewcommand{\thecorollary}{\Alph{appendixc}.\arabic{corollary}}
        \renewcommand{\theequation}{\Alph{appendixc}.\arabic{equation}}
        \noindent{\tenbf Appendix \theappendixc #1}\par\vspace{5pt}}
\newcommand{\subappendix}[1] {\vspace{12pt}
        \refstepcounter{subappendixc}
        \noindent{\bf Appendix \thesubappendixc. {\kern1pt \bfit #1}}
	\par\vspace{5pt}}
\newcommand{\subsubappendix}[1] {\vspace{12pt}
        \refstepcounter{subsubappendixc}
        \noindent{\rm Appendix \thesubsubappendixc. {\kern1pt \tenit #1}}
	\par\vspace{5pt}}

\newcommand{\textlineskip}{\baselineskip=13pt}
\newcommand{\smalllineskip}{\baselineskip=10pt}

\def\eightcirc{
\begin{picture}(0,0)
\put(4.4,1.8){\circle{6.5}}
\end{picture}}
\def\eightcopyright{\eightcirc\kern2.7pt\hbox{\eightrm c}} 

\newcommand{\copyrightheading}[1]
	{\vspace*{-2.5cm}\smalllineskip{\flushleft
	{\footnotesize International Journal of Modern Physics B, #1}\\
	{\footnotesize $\eightcopyright$\, World Scientific Publishing
	 Company}\\
	 }}

\newcommand{\pub}[1]{{\begin{center}\footnotesize\smalllineskip 
	Received #1\\
	\end{center}
	}}

\def\abstracts#1#2#3{{
	\centering{\begin{minipage}{4.5in}\baselineskip=10pt\footnotesize
	\parindent=0pt #1\par 
	\parindent=15pt #2\par
	\parindent=15pt #3
	\end{minipage}}\par}} 

\renewenvironment{thebibliography}[1]			
	{\frenchspacing
	 \ninerm\baselineskip=11pt
	 \begin{list}{\arabic{enumi}.}
	{\usecounter{enumi}\setlength{\parsep}{0pt}
	 \setlength{\leftmargin 12.7pt}{\rightmargin 0pt} 
	 \setlength{\itemsep}{0pt} \settowidth
	{\labelwidth}{#1.}\sloppy}}{\end{list}}

\newcounter{itemlistc}
\newcounter{romanlistc}
\newcounter{alphlistc}
\newcounter{arabiclistc}

\newcommand{\fcaption}[1]{
        \refstepcounter{figure}
        \setbox\@tempboxa = \hbox{\footnotesize Fig.~\thefigure. #1}
        \ifdim \wd\@tempboxa > 5in
           {\begin{center}
        \parbox{5in}{\footnotesize\smalllineskip Fig.~\thefigure. #1}
            \end{center}}
        \else
             {\begin{center}
             {\footnotesize Fig.~\thefigure. #1}
              \end{center}}
        \fi}

\newcommand{\tcaption}[1]{
        \refstepcounter{table}
        \setbox\@tempboxa = \hbox{\footnotesize Table~\thetable. #1}
        \ifdim \wd\@tempboxa > 5in
           {\begin{center}
        \parbox{5in}{\footnotesize\smalllineskip Table~\thetable. #1}
            \end{center}}
        \else
             {\begin{center}
             {\footnotesize Table~\thetable. #1}
              \end{center}}
        \fi}

\def\@citex[#1]#2{\if@filesw\immediate\write\@auxout
	{\string\citation{#2}}\fi
\def\@citea{}\@cite{\@for\@citeb:=#2\do
	{\@citea\def\@citea{,}\@ifundefined
	{b@\@citeb}{{\bf ?}\@warning
	{Citation `\@citeb' on page \thepage \space undefined}}
	{\csname b@\@citeb\endcsname}}}{#1}}

\newif\if@cghi
\def\cite{\@cghitrue\@ifnextchar [{\@tempswatrue
	\@citex}{\@tempswafalse\@citex[]}}
\def\citelow{\@cghifalse\@ifnextchar [{\@tempswatrue
	\@citex}{\@tempswafalse\@citex[]}}
\def\@cite#1#2{{$\null^{#1}$\if@tempswa\typeout
	{IJCGA warning: optional citation argument 
	ignored: `#2'} \fi}}

\def\pmb#1{\setbox0=\hbox{#1}
	\kern-.025em\copy0\kern-\wd0
	\kern.05em\copy0\kern-\wd0
	\kern-.025em\raise.0433em\box0}

\def\fnt#1#2{\footnotetext{\kern-.3em
	{$^{\mbox{\scriptsize #1}}$}{#2}}}

\def\fpage#1{\begingroup
\voffset=.3in
\thispagestyle{empty}\begin{table}[b]\centerline{\footnotesize #1}
	\end{table}\endgroup}

\def\runninghead#1#2{\pagestyle{myheadings}
\markboth{{\protect\footnotesize\it{\quad #1}}\hfill}
{\hfill{\protect\footnotesize\it{#2\quad}}}}
\font\tenrm=cmr10
\font\tenit=cmti10 
\font\tenbf=cmbx10
\font\bfit=cmbxti10 at 10pt
\font\ninerm=cmr9

\font\eightrm=cmr8

\def\qed{\hbox{${\vcenter{\vbox{			
   \hrule height 0.4pt\hbox{\vrule width 0.4pt height 6pt
   \kern5pt\vrule width 0.4pt}\hrule height 0.4pt}}}$}}

\def\bsc{{\sc a\kern-6.4pt\sc a\kern-6.4pt\sc a}}	
\def\bflatex{\bf L\kern-.30em\raise.3ex\hbox{\bsc}\kern-.14em 
T\kern-.1667em\lower.7ex\hbox{E}\kern-.125em X} 

\begin{document}

\runninghead{
Vertex corrections in  gauge theories for 
two-dimensional condensed matter systems} {
Vertex corrections in  gauge theories for 
two-dimensional condensed matter systems}

\normalsize\textlineskip
\thispagestyle{empty}
\setcounter{page}{1}

\copyrightheading{}			

\vspace*{0.88truein}

\fpage{1}
\centerline{\bf VERTEX CORRECTIONS IN GAUGE THEORIES FOR}
\vspace*{0.035truein}
\centerline{\bf TWO-DIMENSIONAL CONDENSED MATTER SYSTEMS}
\vspace*{0.37truein}
\centerline{\footnotesize PETER KOPIETZ}
\vspace*{0.015truein}
\centerline{\footnotesize\it 
Department of Physics and Astronomy, University of California,}
\baselineskip=10pt
\centerline{
\footnotesize\it 
 Los Angeles, CA 90095, USA\cite{address}}
\baselineskip=10pt
\centerline{\footnotesize\it and}
\baselineskip=10pt
\centerline{\footnotesize\it 
Institut f\"{u}r Theoretische Physik der Universit\"{a}t G\"{o}ttingen, Bunsenstrasse 9,}
\baselineskip=10pt
\centerline{
\footnotesize\it 
D-37073 G\"{o}ttingen, Germany}

\vspace*{10pt}
\pub{December 1997}

\vspace*{0.21truein}
\abstracts{
We calculate the  self-energy of two-dimensional
fermions that are coupled to transverse gauge fields, taking two-loop
corrections into account.
Given a bare gauge field propagator
that diverges for small momentum transfers
$q$ as $1 / q^{\eta}$, $1 < \eta \leq 2$,
the fermionic self-energy
without vertex corrections vanishes 
for small frequencies $\omega$ as
$\Sigma ( \omega ) \propto \omega^{\gamma}$ with
$\gamma = {\frac{2}{1 + \eta}} < 1$. We show that 
inclusion of the leading radiative correction
to the fermion - gauge field vertex 
leads to
$\Sigma ( \omega ) \propto \omega^{\gamma}
 [ 1 - a_{\eta} \ln ( \omega_0 / \omega ) ]$,
where $a_{\eta}$ is a positive numerical constant and
$\omega_0$ is some finite energy scale.
The negative logarithmic correction 
is consistent with the scenario
that higher order vertex corrections 
push the exponent $\gamma$ to larger values.
}{}{}

\vspace*{1pt}\textlineskip	
\section{Introduction}	
\vspace*{-0.5pt}
\noindent
The problem of two-dimensional fermions that are coupled to
a transverse gauge field has recently 
received a lot of attention, because it arises 
as effective low-energy theory in two
different physical contexts.
On the one hand, the infrared physics of the two-dimensional
$t-J$-model has been argued to be correctly described by 
fermions and bosons that are coupled to an Abelian gauge 
field\cite{Baskaran88,Ioffe89,Lee89}.
The other example are half-filled quantum Hall 
systems\cite{Halperin93},
where a fictitious Chern-Simons gauge field can be used
to attach two quanta of a magnetic flux to the 
physical electron, thus forming a new (presumably stable)
quasi-particle, the so-called composite fermion\cite{Jain89}.
Within mean field theory, where fluctuations of the
gauge field are ignored, the fermions are
assumed to form a conventional Fermi liquid.
Many authors have considered the stability of the
Fermi liquid with respect to fluctuations of the gauge field.
When one calculates the fermionic self-energy $\Sigma ( \omega )$ 
to first order in the dynamically screened propagator
of the gauge field, one finds for small 
frequencies\cite{Halperin93,Blok93,Ioffe94,Altshuler94}
 \begin{equation}
 \Sigma ( \omega ) \propto \omega^{\gamma} \;  ,
 \label{eq:sigmaNFL}
 \end{equation}
with $\gamma < 1$. This implies a vanishing quasi-particle residue
 $ Z = \lim_{\omega \rightarrow 0}
[ 1 - \partial \Sigma ( \omega ) / \partial \omega ]^{-1}$, 
so that the system cannot be a Fermi liquid.
Because first order perturbation theory qualitatively
changes the analytic structure of the non-interacting Green's function and
there is no obvious small parameter in the problem, it
is necessary to address the effect of higher orders in perturbation
theory.

Clearly, in order to understand the physics of gauge fields 
in these strongly correlated systems, conventional perturbative many-body
theory is not sufficient. At least one should
sum properly chosen infinite sub-classes of Feynman diagrams.
Over the past few years a variety of methods
have been proposed to resum the most divergent terms to all orders in
perturbation theory\cite{Ioffe94,Altshuler94,Khveshchenko93,Gan93,%
Plochinski94,Nayak94,%
Kwon94,Onoda95,Chakravarty95,Kopietz96a,Kopietz97}.
However, so far a general agreement as not been achieved.  
In particular, in Ref.\cite{Kopietz97} it was found
by means of a  non-perturbative functional 
integral approach\cite{Kopietz96,Kopietzbook}
that higher orders
completely change the scenario suggested by 
a one-loop calculation: the non-analyticities suggested by
lowest order perturbation theory were found to be partially removed, 
so that the spectral function exhibits a well-defined
quasi-particle peak.
This implies that there must exist some
higher order terms 
in the perturbative expansion of the self-energy
that are more singular than the one-loop result.
In this work we shall explicitly calculate the
two-loop correction to the self-energy due to the
leading radiative correction to the fermion - gauge field vertex,
and show that it contains an additional factor proportional
to
$\ln ( \omega_0 / \omega )$ as compared with the one-loop result
($\omega_0$ is some finite frequency, see 
Eq.(\ref{eq:selfresfinal}) below).
For reasons that will be explained in detail in Sec.3,
this logarithmic correction has been missed in 
a previous calculation by Altshuler, Ioffe and Millis\cite{Altshuler94}.
The existence of a logarithmically divergent correction
implies
that the true infrared behavior of the
fermionic self-energy cannot be obtained from 
a one-loop calculation.
In fact, the prefactor of the logarithm turns out
to be negative (relative to the one-loop self-energy, see
Eq.(\ref{eq:selfresfinal})), so that
it is consistent with the 
scenario that the summation of vertex corrections to infinite orders
leads to an {\it{increase}} 
of the exponent $\gamma$ in Eq.(\ref{eq:sigmaNFL}), and possibly 
restores Fermi liquid behavior\cite{Kopietz97} 
(which requires $\gamma \geq 1$).

\section{The self-energy without vertex corrections}
\label{sec:theself}
\begin{figure}
\epsfysize4cm 
\hspace{23mm}
\epsfbox{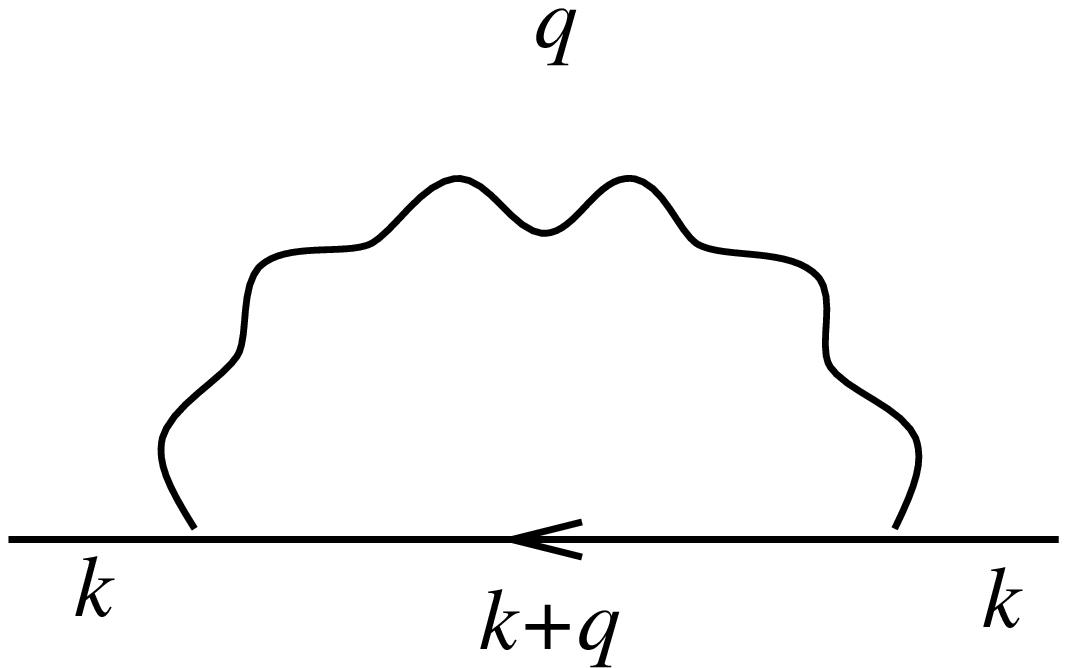}
\fcaption{
Leading fluctuation correction to the fermionic self-energy.
The solid arrow is the mean-field Green's function, and the
wavy-line is the RPA screened propagator of the
gauge field, see Eqs.(\ref{eq:G0def}) and (\ref{eq:hrpa}). }
\label{fig:rainbow}
\end{figure}

\begin{figure}
\epsfysize4cm 
\hspace{23mm}
\epsfbox{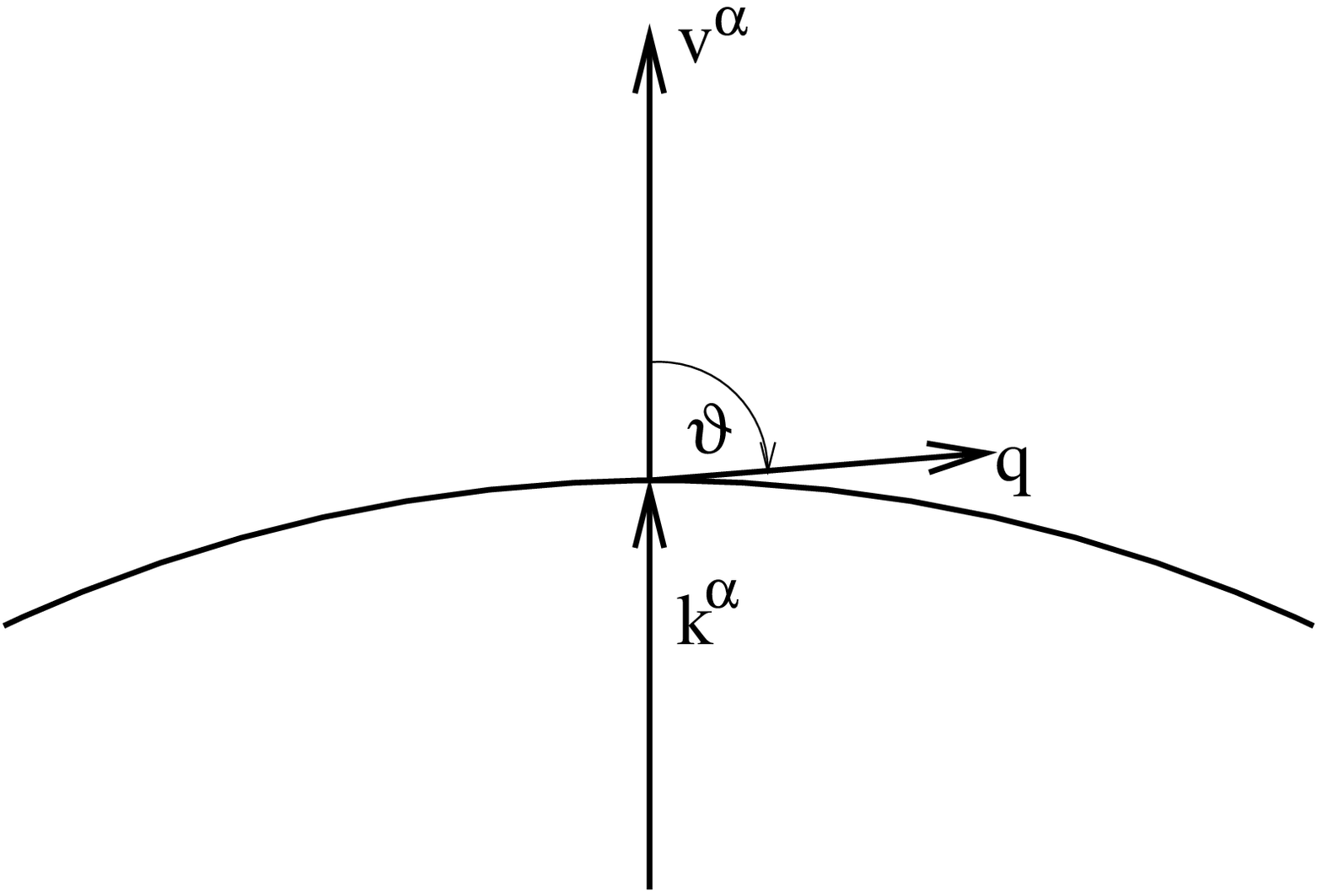}
\fcaption{Definition of circular coordinates
centered at point ${\bf{k}}^{\alpha}$ on the Fermi surface.}
\label{fig:coordinate}
\end{figure}

\noindent
Let us begin with a
careful discussion of the 
one-loop fermionic self-energy correction $\Sigma_1$ due to
fluctuations of the gauge field. The
relevant Feynman diagram is shown in Fig.\ref{fig:rainbow}.
Although this diagram has been calculated previously by many
authors\cite{Halperin93,Blok93,Ioffe94,Altshuler94,Onoda95}
let us calculate it once more, using a particular
coordinate system where all wave-vectors are measured 
relative to a fixed point
on the Fermi surface.  The insights gained from this
calculation will be useful for the more difficult
two-loop calculation.
Using the Matsubara formalism,
Fig.\ref{fig:rainbow} represents the following 
expression for the fermionic self-energy,
\begin{equation}
 \Sigma_{1}^{\alpha} ( \tilde{k} ) =
 - \frac{1}{\beta V} \sum_{ \tilde{q} }
 h^{{\rm RPA}, \alpha}_{\tilde{q}} G_0^{\alpha} ( \tilde{k} + \tilde{q} )
 \label{eq:GW}
 \;  ,
 \end{equation}
where $V$ is the volume of the system, $\beta$ is the inverse 
temperature, and we have defined collective labels
$\tilde{k} = [ {\bf{k}} , i \tilde{\omega}_n ]$,
$\tilde{q} = [ {\bf{q}}, i \omega_n]$,
where $\tilde{\omega}_n = 2 \pi (n +  
\frac{1}{2}) / \beta $ are fermionic 
Matsubara frequencies, and
${\omega}_n = 2 \pi n / \beta$ are bosonic ones.
The superscript $\alpha$ indicates that all wave-vectors are measured with
respect to a point ${\bf{k}}^{\alpha}$ 
on the Fermi surface, as shown in Fig.\ref{fig:coordinate}.
The free Green's function is
 \begin{equation}
 G_0^{\alpha} ( \tilde{k} ) = \frac{1}{ i \tilde{\omega}_n - 
 \xi_{ {\bf{k}}^{\alpha} + {\bf{k}}}  }
 \label{eq:G0def}
 \; ,
 \end{equation}
where
 $
 \xi_{ {\bf{k}}^{\alpha} + {\bf{k}} } = {\bf{v}}^{\alpha} \cdot {\bf{k}}
 + { {\bf{k}}^2}/ ({2 m^{\alpha}})
 $
is the mean field
energy dispersion (measured relative to the chemical potential)
for wave-vectors ${\bf{k}}$ close to ${\bf{k}}^{\alpha}$.
Here
${\bf{v}}^{\alpha}$ is the local Fermi velocity 
with magnitude $v_F$,
and $m^{\alpha}$ is the local effective mass
close to
${\bf{k}}^{\alpha}$.  Note that we are not assuming that the Fermi surface
is spherically symmetric; in particular, for $m^{\alpha} \rightarrow
\infty$  the Fermi surface becomes locally flat.
The dynamically screened gauge field propagator
is within the random phase approximation (RPA) and for 
frequencies in the regime
 $| \omega_n | 
 { \raisebox{-0.5ex}{$\; \stackrel{<}{\sim} \;$}}
 v_F q $ given by
 \begin{equation}
 h^{{\rm RPA}, \alpha }_{\tilde{q}}
 = -  \frac{2 \pi  }{m^{\ast}  } \;
  [1 - ({\hat{\bf{v}}^{\alpha}} \cdot {\hat{\bf{q}}} )^2] 
 \frac{v_F q }
 { \Gamma_q  + 
  | \omega_{n}|  
  }
 \label{eq:hrpa}
 \;  ,
 \end{equation}
where we have used the Coulomb gauge, $\hat{\bf{v}}^{\alpha}$
is a unit vector parallel to ${\bf{v}}^{\alpha}$,
and the energy scale $\Gamma_q$ is 
 \begin{equation}
 \Gamma_q = v_F q
 \left( q / q_c  \right)^{\eta } = v_F q_c 
\left( q /{ q_c } \right)^{1 + \eta }
 \; .
 \label{eq:energyscale}
 \end{equation}
The mass $m^{\ast}$ is some effective mass such that
$m^{\ast} v_F \equiv k_{F}$ is a measure for the
average curvature of the Fermi surface.
In contrast, $ m^{\alpha} v_F \equiv k_c $ measures
the {\it{local}} curvature of the Fermi surface.
Throughout this work we shall assume that the momentum
scale $q_{c}$ in Eq.(\ref{eq:energyscale})
is small compared with $k_{F} $ and $k_{c}$, and that
the exponent $\eta$  
is in the interval $1 < \eta \leq 2$. The case $\eta = 2$ 
is relevant to gauge theories of high-temperature superconductors,
as well as half-filled quantum Hall systems with short-range 
density-density interactions.

To perform the integration over the momentum-transfer
${\bf{q}}$ in Eq.(\ref{eq:GW}), we choose the
circular coordinates shown
in Fig.\ref{fig:coordinate}.
For simplicity, we shall  restrict ourselves 
to external wave-vectors of the form
$ {\bf{k}} = k_{\bf{\|}} \hat{\bf{v}}^{\alpha}$, so that
 \begin{equation}
 \xi_{ {\bf{k}}^{\alpha} + {\bf{k}} + {\bf{q}} } =
 \xi_{ k_{\|}   }
 +  ( 1 + \frac{k_{\|}}{k_c} ) v_F q \cos \vartheta
 + \frac{ q^2 }{2 m^{\alpha}}
 \; ,
 \end{equation}
where
 $ \xi_{  k_{\|} }
 = v_F k_{\|} +  k_{\|}^2/(2 m^{\alpha})$.
Keeping in mind that the form (\ref{eq:hrpa}) for the gauge
field propagator is valid for $| \omega_n | 
{ \raisebox{-0.5ex}{$\; \stackrel{<}{\sim} \;$}}
 v_F q $, 
and imposing an ultraviolet cutoff $\kappa$ 
on the $q$-integration
in Eq.(\ref{eq:GW}) (anticipating that the 
leading behavior for small frequencies is dominated by the
infrared singularities of the integrand, we may choose
$q_c \ll \kappa \ll k_F$) 
we obtain from
Eq.(\ref{eq:GW}) for $V \rightarrow \infty$ and
$\beta \rightarrow \infty$,
 \begin{eqnarray}
 \Sigma_{1}^{\alpha} ( k_{\|} , i \tilde{\omega}_n )
 & = & 
 \frac{ 1}{(2 \pi)^2 m^{\ast} }
 \int_{0}^{\kappa} dq  q
 \int_{ - v_F q}^{v_F q} d \omega \frac{v_F q }{ | \omega | + \Gamma_q}
 \nonumber
 \\
 &   &  \hspace{-20mm} \times
 \int_{- \pi}^{ \pi} 
 d \vartheta
 \frac{ \sin^2 \vartheta}{
 G_0^{-1} + i \omega -  
 (  1 + \frac{k_{\|}}{k_c} ) v_F q \cos \vartheta - 
\frac{ q^2}{2 m^{\alpha} }  }
 \label{eq:sigma1int}
 \; ,
 \end{eqnarray}
with $G_0^{-1} = i \tilde{\omega}_n - \xi_{k_{\|}}$.
Scaling out a factor of $v_F q (1 + \frac{k_{\|}}{k_c} )$
and defining the complex variable
\begin{equation}
 W ( q , \omega ) = 
  \frac{ G_0^{-1} + i \omega - \frac{ q^2}{2 m^{\alpha} }}{v_F q 
 ( 1 + \frac{k_{\|}}{k_c} ) }
 \label{eq:Wdef}
 \; ,
 \end{equation}
Eq.(\ref{eq:sigma1int}) can also be written as
 \begin{equation}
 \Sigma_{1}^{\alpha} ( k_{\|} , i \tilde{\omega}_n )
  =  
\frac{ 1}{2 \pi m^{\ast} ( 1 + \frac{k_{\|} }{k_c})}
 \int_{0}^{\kappa} dq  q
 \int_{ - v_F q}^{v_F q} d \omega 
 \frac{ 
 Z ( W ( q , \omega ))
}{ | \omega | + \Gamma_q}
 \label{eq:sigma2int}
 \; .
 \end{equation}
Here the complex function $Z ( W)$ is defined by\cite{Henrici88}
 \begin{equation}
 Z ( W ) = \frac{1}{\pi} \int_{0}^{\pi} d \vartheta
 \frac{\sin^2 \vartheta}{ W - \cos \vartheta}
 = W - \sqrt{ W^2 -1 }
 \; ,
 \label{eq:Zdef}
 \end{equation}
where the root has to be taken such that $| Z | < 1$.
It turns out that the leading infrared behavior of
Eq.(\ref{eq:sigma2int}) is due to the regime where
$|W ( q , \omega ) | \ll 1$. To see this,
note that
 \begin{equation}
 Z ( W ) = 
 \left\{ \begin{array}{ll}
 - i {\rm sgn} ( {\rm  Im } W ) + W + O ( W^2)
 & \mbox{if $| W | \ll 1$}
 \\
 & \\
 ( 2 W )^{-1} + O ( W^{-2} )
 & \mbox{if $ | W | \gg 1$}
 \end{array}
 \right.
 \;  .
 \label{eq:Wasym}
 \end{equation}
Because $ |\omega | / ( v_F q ) 
{ \raisebox{-0.5ex}{$\; \stackrel{<}{\sim} \;$}} 1$ 
and $\frac{q^2 }{  2 m^{\alpha} } 
{ \raisebox{-0.5ex}{$\; \stackrel{<}{\sim} \;$}} v_F q$ 
in the domain of integration in Eq.(\ref{eq:sigma2int}), the
condition $| W | \ll 1$ is equivalent with
 \begin{equation}
q { \raisebox{-0.5ex}{$\; \stackrel{>}{\sim} \;$}} k_0  
\equiv  {\rm max} 
\left\{ | k_{\|} | , | \tilde{\omega}_n | / v_F \right\}
 \label{eq:k0def}
 \; .
 \end{equation}
To evaluate the integral in Eq.(\ref{eq:sigma2int}), we
subdivide the $q$-integration into the regimes
$0 < q < k_0$ and $k_0 < q < \kappa$, and use the corresponding
asymptotic forms of $Z (W )$ 
given in Eq.(\ref{eq:Wasym}).
Using the fact that $| Z (W ) | \leq 1$,
it is not difficult to show that  the contribution
from the regime $0 < q  < k_0$  is 
 \begin{equation}
 \Sigma_{\rm 1}^{\alpha} ( k_{\|} , i \tilde{\omega}_n )_{( q < k_0)}
 \propto \frac{ k_0^2}{m^{\ast}} \ln (  q_c / k_0 )
 \label{eq:SigmaGWsmall}
 \; .
 \end{equation}
Obviously,  for small $\tilde{\omega}_n$ and $k_{\|}$ this
correction is negligible.
On the other hand, in
the regime $k_0 < q < \kappa$ we may approximate
(see Eqs.(\ref{eq:Wdef}) and (\ref{eq:Wasym}))
 \begin{equation}
 Z ( W ( q , \omega ) ) \approx - i {\rm sgn} (  \tilde{\omega}_n + \omega )
 \label{eq:ZWsmall}
 \; .
 \end{equation}
The corresponding contribution to the self-energy is
for $| k_{\|} | \ll k_c$
 \begin{equation}
 \Sigma_{1}^{\alpha} ( k_{\|} , i \tilde{\omega}_n )_{( q > k_0)}
  =  \frac{ - i }{2 \pi m^{\ast} }
 \int_{k_0}^{\kappa} dq  q
 \int_{ - v_F q}^{v_F q} d \omega
 \frac{
  {\rm sgn} ( \tilde{\omega}_n + \omega )
 }{ | \omega | + \Gamma_q}  
 \label{eq:GWspherical3}
 \;  .
 \end{equation}
Using the fact that the rest of the integrand is an even function
of $\omega$, 
we may replace under the integral sign
 \begin{equation}
 {\rm sgn} ( \tilde{\omega}_n + \omega )
  \rightarrow 
 {\rm sgn} ( \tilde{\omega}_n + \omega ) - 
 {\rm sgn} ( \omega ) 
  = 
  2 {\rm sgn} ( \tilde{\omega}_n )   
 \Theta ( - \omega ( \tilde{\omega}_n + \omega ) )
 \label{eq:sgnreplace}
 \;  .
 \end{equation}
Then it is easy to show that
 \begin{eqnarray}
 \Sigma_{\rm 1}^{\alpha} ( k_{\|} , i \tilde{\omega}_n )_{( q > k_0)}
 & = & 
 \frac{ - i {\rm sgn} ( \tilde{\omega}_n ) }{ \pi m^{\ast} }
 \int_{k_0}^{\kappa} dq   q
 \ln \left( 1 + \frac{ | \tilde{\omega}_n |}{ \Gamma_q } \right)
 \label{eq:GWspherical4}
 \; .
 \end{eqnarray}
Introducing the new integration variable 
 $y = {\Gamma_q } / { | \tilde{\omega}_n | }$
and dimensionless frequencies 
 $\bar{\omega}_n = { \tilde{\omega}_n }/ ({ v_F q_c })$ and
wave-vectors
 $\bar{k}_{\|} = { k_{\|}} / {q_c} $,
we have
 \begin{equation}
 q = q_c | \bar{\omega}_n y |^{\frac{1}{ 1 + \eta}}
 \; \; \; , \; \; \; 
 dq = q_c | \bar{\omega}_n |^{\frac{1}{ 1 + \eta}}
 \frac{y^{\frac{- \eta  }{1 + \eta} }}{ 1 + \eta} d y
  \; ,
 \label{eq:qy}
 \end{equation}
so that
 \begin{equation}
 \frac{\Sigma_{\rm 1}^{\alpha} ( k_{\|} , i \tilde{\omega}_n )_{( q > k_0)}}{v_F q_c}
  =  
  -    i  g  \; {\rm sgn} ( \tilde{\omega}_n )
  | \bar{\omega}_n |^{ \frac{2}{1 + \eta} }
 F_{\rm 1 } ( \bar{k}_{\|}   , i \bar{\omega}_n )  
 \label{eq:GWspherical5}
 \; ,
 \end{equation}
where $g = q_c / k_F \ll 1 $ is a dimensionless coupling constant.
The dimensionless function
 $F_{\rm 1 } ( \bar{k}_{\|}   , i \bar{\omega}_n ) $ is  defined by
 \begin{equation}
 F_{\rm 1 } ( \bar{k}_{\|}   , i \bar{\omega}_n ) 
 = \frac{1}{\pi ( 1 + \eta )} \int_{y_0}^{y_1} dy y^{\frac{1 - \eta}{1 + \eta}} \ln ( 1 + 1 / y )
 \label{eq:cetadef}
 \;  ,
 \end{equation}
with the lower limit 
 \begin{equation}
 y_0 = \frac{\Gamma_{k_0}}{ | \tilde{\omega}_n | } =
 \left( \frac{k_0}{q_c} \right)^{\eta} \frac{ v_F k_0}{ | \tilde{\omega}_n | }
 = \left\{
 \begin{array}{ll} | \bar{\omega}_n |^{\eta} 
 & \mbox{if $ | \bar{\omega}_n | \geq 
 | \bar{k}_{\|} | $} \\
 &  \\
 | \bar{k}_{\|} |^{1 + \eta} | \bar{\omega}_n |^{-1}
 & \mbox{if $ | \bar{\omega}_n | \ll
  | \bar{k}_{\|} | $} 
 \end{array}
 \right.
 \label{eq:y0def}
 \; , 
 \end{equation}
and the upper limit 
  \begin{equation}
  y_1 = \frac{\Gamma_{\kappa}}{ | \tilde{\omega}_n | } = \frac{1}{ | \bar{\omega}_n | }
  \left( \frac{\kappa}{q_c } \right)^{1 + \eta} 
  \; .
  \label{eq:y1def}
  \end{equation}
Note that for large $y$ the integrand in 
Eq.(\ref{eq:cetadef}) vanishes as
$y^{\frac{ 1 - \eta}{ 1 + \eta} - 1}$, so that for
$\eta > 1$ the integral
is ultraviolet convergent. 
To obtain the limiting behavior of the self-energy for $|\bar{\omega}_n | \rightarrow 0$,
we may let $y_1 \rightarrow \infty$.
Furthermore, for $| \bar{\omega}_n | \geq  | \bar{k}_{\|} |$ we may
set $y_0 =0$ in the lower limit of Eq.(\ref{eq:cetadef}), 
so that
 \begin{equation}
 F_{\rm 1} ( \bar{k}_{\|} , i \bar{\omega}_n)  \approx
 \frac{1}{\pi ( 1 + \eta )} \int_{0}^{\infty} dy 
 y^{\frac{1 - \eta}{1 + \eta}} \ln ( 1 + 1 / y )
 \equiv c_{\eta}
 \; .
 \label{eq:FGWconst}
 \end{equation}
Hence, to leading order
 \begin{equation}
  \frac{ \Sigma_{1}^{\alpha} ( k_{\|} , i \tilde{\omega}_n )_{( q > k_0)}}{v_F q_c}
  \sim  
  -    
  i c_{\eta} g \; 
  {\rm sgn} ( \tilde{\omega}_n ) 
  | \bar{\omega}_n |^{ \frac{2}{1 + \eta} }
 \; \; ,
 \; \; | \bar{\omega}_n | \geq  | \bar{k}_{\|} |
 \label{eq:GWspherical6}
 \; .
 \end{equation} 
For completeness, let us also discuss the regime
$ | \bar{\omega}_n | \ll  | \bar{k}_{\|} |$, where
$y_0 \gg 1$. Then
 \begin{equation}
 F_{1} ( \bar{k}_{\|} ,
 i \bar{\omega}_n ) = \frac{1 + \eta}{\eta - 1} y_0^{-1} + O ( y_0^{-2} )
 \label{eq:Cetalarge}
 \; ,
 \end{equation}
so that
 \begin{equation}
 \frac{\Sigma_{1}^{\alpha} ( k_{\|} , 
 i \tilde{\omega}_n )_{( q > k_0)}}{v_F q_c}
 \sim 
  - 
 \frac{   g  }{\pi ( \eta -1 ) }
  \frac{  i \bar{\omega}_n 
  | \bar{\omega}_n |^{ \frac{2}{1 + \eta} }}{
  | \bar{k}_{\|}|^{1+ \eta}  }
 \; \;  , \; \;  | \bar{\omega}_n | \ll | \bar{k}_{\|} |
 \; .
 \end{equation}
The spectral function can now be obtained from the 
retarded self-energy
  $\Sigma_{1}^{\alpha} ( k_{\|} , \omega + i 0^{+})$,
\begin{equation}
 A ( k_{\|} , \omega  )
 = - \frac{1}{\pi} {\rm Im}
 \left[ \frac{1}{ \omega - \xi_{k_{\|}} -
  \Sigma_{1}^{\alpha} ( k_{\|} , \omega + i 0^{+}) }
 \right]
 \label{eq:AGWret}
 \; .
 \end{equation}
To perform the analytic continuation
(i.e.  $
 i \tilde{\omega}_n \rightarrow  \omega + i 0^{+}$,
 ${\rm sgn} ( \tilde{\omega}_n ) \rightarrow 1$, and
 $| \tilde{\omega}_n | \rightarrow - i \omega$)
we write
 \begin{equation}
 i {\rm sgn} ( \tilde{\omega}_n ) 
  | \tilde{\omega}_n |^{ \frac{2}{1 + \eta} }
  = 
 i \tilde{\omega}_n  
 \left[ - i 
 {\rm  sgn} ( \tilde{\omega}_n ) 
 i \tilde{\omega}_n \right]^{\frac{1 - \eta}{1 + \eta}}
 \label{eq:iomegaomega}
 \;  ,
 \end{equation}
so that 
 \begin{eqnarray}
 i {\rm sgn} ( \tilde{\omega}_n ) 
  | \tilde{\omega}_n |^{ \frac{2}{1 + \eta} }
  & \rightarrow & 
  {\rm sgn} ( \omega ) | \omega |^{\frac{2}{1 + \eta}} \exp \left[ i 
  {\rm sgn}{ ( \omega )} 
  \frac{ \pi}{2}
  \frac{ \eta -1}{\eta + 1} 
  \right]
  \nonumber
  \\
  & = &  
  | \omega |^{\frac{2}{1 + \eta}}  
  \left[ 
  \lambda^{\prime} 
  {\rm sgn} ( \omega )  
  + i 
  \lambda^{\prime \prime}  \right]
  \label{eq:analyticcont2}
  \; ,
  \end{eqnarray}
with
  $ \lambda^{\prime} = \cos ( \frac{\pi}{2} \frac{ \eta -1}{\eta + 1}
 )$ and
  $\lambda^{\prime \prime } = \sin ( \frac{\pi}{2} 
 \frac{ \eta -1}{\eta + 1} )$.
Note that for $\eta > 1$ both $\lambda^{\prime}$ and 
$\lambda^{\prime \prime}$ are positive real constants.
We conclude that 
the retarded self-energy is for small frequencies in the regime
$| \bar{\omega} | \geq | \bar{k}_{\|}|$ given by
 \begin{equation}
  \frac{ \Sigma_{ 1}^{\alpha} ( k_{\|} , \omega + i 0^{+})}{v_F q_c}
  \sim  
  -    c_{\eta} g
  | \bar{\omega} |^{ \frac{2}{1 + \eta} }
  \left[ 
  \lambda^{\prime} 
  {\rm sgn} ( \omega )  
  + i 
  \lambda^{\prime \prime}  \right]
  \label{eq:sigmaGWret}
  \;  .
  \end{equation}
Note that
${\rm Im} \Sigma_{1}^{\alpha} ( k_{\|} , \omega + i 0^{+} ) < 0$, 
so that the retarded Green's function
is analytic in the entire upper half of the
complex frequency plane (as it should), and the
corresponding spectral function (\ref{eq:AGWret})
is positive (as it should).
However,
the spectral function does not exhibit a well-defined
quasi-particle peak,
because according to Eq.(\ref{eq:sigmaGWret})
the imaginary part of the self-energy has the same order of magnitude as the
real part.
Thus, lowest order perturbation theory suggests that
fluctuations of the gauge field  completely destroy the Fermi liquid
state predicted by mean field theory.

From our rather detailed derivation 
it is now easy to identify the regions in energy-momentum space
that are responsible for this non-Fermi liquid behavior.
Obviously, in deriving Eq.(\ref{eq:sigmaGWret}) we may
restrict ourselves to the regime where
$| W ( q , \omega ) |$ is small compared with unity,
which requires $k_{0} { \raisebox{-0.5ex}{$\; \stackrel{<}{\sim} \;$}}
q$.
Moreover, the leading term in the expansion of the function
$Z ( W )$ (see Eq.(\ref{eq:ZWsmall})) for small
$| W |$ is due to angles $\vartheta$ 
in the integral representation (\ref{eq:Zdef}) of this function
such that
$| \cos \vartheta |
{ \raisebox{-0.5ex}{$\; \stackrel{<}{\sim} \;$}} | W | \ll 1$.
This is most easily seen by shifting
$x = \vartheta - \pi /2$ in Eq.(\ref{eq:Zdef}), and using the fact
that for small $|W|$  the leading behavior of the integral
is determined by $|x - {\rm Re}{W} | 
{ \raisebox{-0.5ex}{$\; \stackrel{<}{\sim} \;$}}
 | {\rm Im} W | \ll 1$. Hence,
 \begin{equation}
  Z ( W ) \approx - \frac{1}{\pi} \int_{ - \pi /2}^{\pi / 2}
 dx \frac{1}{x - {\rm Re{W}} - i {\rm Im} W}
 \; .
 \label{eq:intsmallx}
 \end{equation}
Using 
 ${\rm Im} ({ x \pm i \epsilon} )^{-1}=  \mp i \pi 
{\rm{ sgn}} ( \epsilon )$,
we see that $
 Z ( W ) \approx  - i {\rm{ sgn Im}} ( W )$ for small 
$| W |$, in agreement with Eq.(\ref{eq:ZWsmall}).

The result (\ref{eq:sigmaGWret})
remains correct if we sum in addition all diagrams 
where the gauge field propagators do not intersect
(thus neglecting vertex corrections).
This is  formally achieved by
replacing
the non-interacting Green's function
$G_{0}^{\alpha} $ on the right-hand side of Eq.(\ref{eq:GW})
by the one-loop renormalized Green's function
$G_{1}^{\alpha}$, thus turning Eq.(\ref{eq:GW}) into
an integral equation. 
Assuming that the momentum-dependence of the
so-defined self-consistent  self-energy
$\tilde{\Sigma}_{1}^{\alpha}$ is again negligible, 
the expression of $\tilde{\Sigma}_{1}^{\alpha}$
is still of the form (\ref{eq:sigma2int}), except that
$W ( q , \omega )$ should be replaced by
\begin{equation}
 \tilde{W }( q , \omega ) = 
  \frac{ G_0^{-1} + i \omega - \tilde{\Sigma}_{1}^{\alpha}
 ( i \tilde{\omega}_n + i \omega ) - \frac{ q^2}{2 m^{\alpha} }}{v_F q 
 ( 1 + \frac{k_{\|}}{k_c} ) }
 \label{eq:Wtildedef}
 \; .
 \end{equation}
Because ${\rm sgn} {\rm Im} ( W ) = {\rm sgn} {\rm Im} ( \tilde{W} )$,
the resulting self-consistent self-energy is again given by
Eq.(\ref{eq:GWspherical3}), but with
the lower limit $k_{0}$ replaced by 
$\tilde{k}_{0} \approx q_c 
| \bar{\omega}_n |^{\frac{2}{1 + \eta}}$,
so that $| \tilde{W} | 
{ \raisebox{-0.5ex}{$\; \stackrel{<}{\sim} \;$}} 1$.
This leads to the new lower
limit
$\tilde{y}_{0} = | \bar{ \omega}_n | $ in Eq.(\ref{eq:cetadef}).
Obviously,
for $| \bar{ \omega}_n | \rightarrow 0$ we may still set 
$\tilde{y}_0 = 0$. Thus, higher order diagrams
without crossings of the gauge field lines
do not modify the result predicted by lowest order
perturbation theory.

Let us summarize what we have learned 
so far. Within a perturbative
calculation to first order in the RPA screened
gauge field propagator, one finds that the fluctuations
of the transverse gauge field give rise to a non-analytic
contribution to the fermionic self-energy, implying 
the non-existence of well defined quasi-particles.
Denoting by ${\bf{q}}$ the momentum transfer and by $\omega$ the energy
transfer mediated by the gauge field, the non-analyticity in the self-energy
arises from energy-momentum transfers in the regime
 \begin{equation}
 | \cos \vartheta | { \raisebox{-0.5ex}{$\; \stackrel{<}{\sim} \;$}} 
 \frac{ | \omega | }{v_F q } 
  \ll
 1 
 \label{eq:regimesmomtrans}
 \; ,
 \end{equation}
where $\vartheta $ is the angle between ${\bf{q}}$ and the local
Fermi velocity ${{\bf{v}}}^{\alpha}$.
In other words, the
non-Fermi liquid behavior is entirely due to momentum transfers
that are almost parallel to the Fermi surface, 
corresponding to $| \cos \vartheta  | \approx | \vartheta \pm 
\frac{\pi}{2} | \ll 1$.
As will be discussed 
in Sec.3.2, this greatly reduces the usefulness of
the well-known Ward-identities
for estimating the importance of vertex corrections.

\section{Vertex corrections}

\subsection{General remarks}
\noindent
Given the fact that the leading fluctuation correction qualitatively 
modifies the
mean field result, one should worry about higher order
corrections. This problem has been addressed previously by
Altshuler, Ioffe, 
and Millis\cite{Altshuler94} (AIM), and by 
Stern and Halperin\cite{Stern95}.
We shall comment on these works below.

\begin{figure}
\epsfysize3.3cm 
\hspace{23mm}
\epsfbox{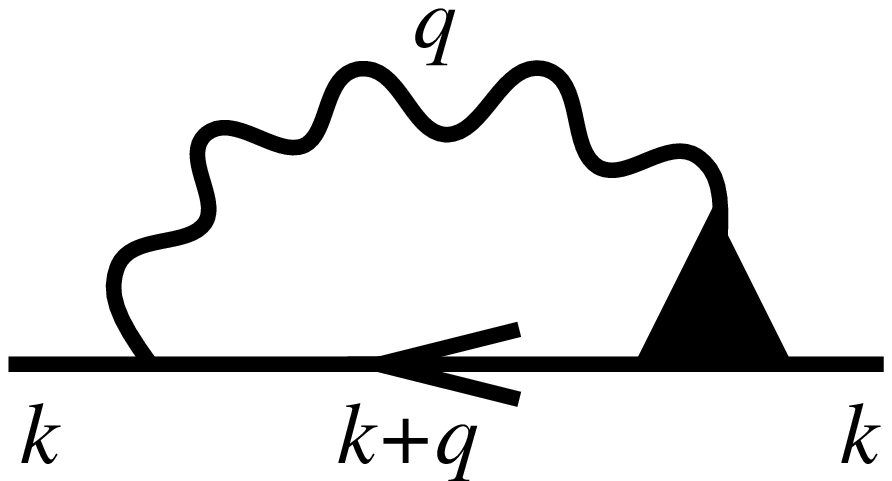}
\fcaption{
Skeleton diagram for the exact self-energy.
The solid arrow is the exact Green's function, the wavy line is the
exact screened gauge field propagator, and the solid triangle represents
the exact three-legged vertex.
}
\label{fig:skeleton}
\end{figure}

\begin{figure}
\epsfysize4.5cm 
\hspace{23mm}
\epsfbox{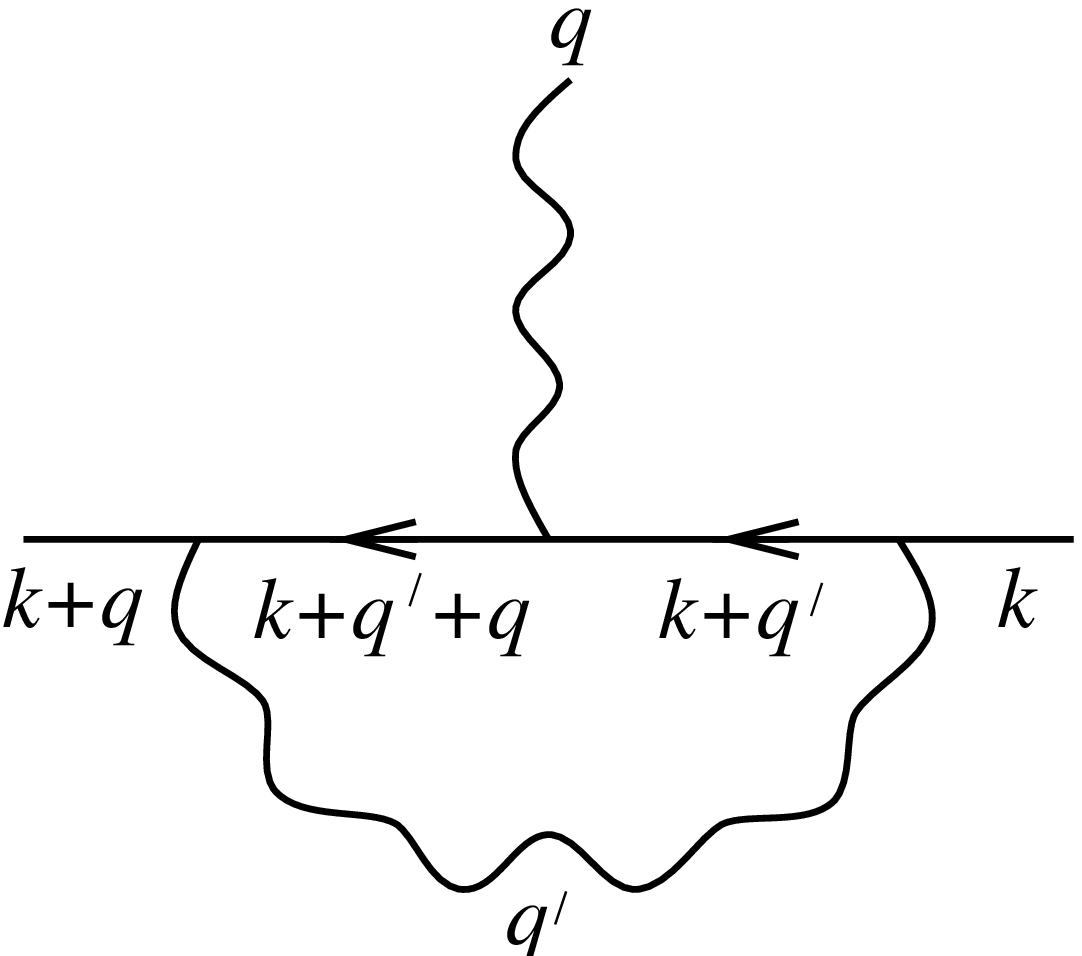}
\fcaption{
Leading radiative correction
to the fermion - gauge field vertex. The notation is the same as in
Fig.\ref{fig:rainbow}.
}
\label{fig:vertex}
\end{figure}

The general structure of the self-energy is conveniently
represented in terms of the skeleton diagram shown in Fig.\ref{fig:skeleton}.
The corresponding analytic expression is
 \begin{equation}
 \Sigma^{\alpha} ( \tilde{k} ) =
 - \frac{1}{\beta V} \sum_{ \tilde{q} }
 h^{ \alpha}_{\tilde{q}} 
 \Lambda^{\alpha} ( \tilde{k} ; \tilde{q} )
 G^{\alpha} ( \tilde{k} + \tilde{q} )
 \label{eq:skeleton}
 \;  ,
 \end{equation}
where $h^{\alpha}_{\tilde{q}}$ is the exact propagator of the gauge field,
$G^{\alpha} ( \tilde{k} )$ is the exact Green's function, and
$\Lambda^{\alpha} ( \tilde{k} , \tilde{q} )$ is the exact 
three-legged vertex.
Obviously, three physically different types of 
corrections can be distinguished. First of all, there are
corrections to the
gauge field propagator beyond the RPA. Because the interaction 
mediated by the
gauge field is most singular for small momentum transfers,
the closed loop theorem discussed in 
Refs.\cite{Kopietzbook,Kopietz95,Metzner97}
guarantees that these corrections are small and can be safely ignored.
This has been explicitly verified at two-loop order by
Kim {\it{et al.}}\cite{Kim94}. The second type of corrections
consists of diagrams without crossings 
of gauge field lines; these contribute to the
exact Green's function $G^{\alpha} ( \tilde{k} + \tilde{q} )$
on the right-hand side of Eq.(\ref{eq:skeleton}). 
At the end of Sec.2 we have shown that 
diagrams of this
type do not modify the leading infrared behavior
of the one-loop self-energy.
However, it is a priori not clear
whether this remains true for higher order diagrams involving
more than one loop.
In fact, according to Refs.\cite{Ioffe94,Altshuler94} 
it is essential to take
these corrections into account by
replacing $G_0 \rightarrow G_1$ in internal loops
of higher order diagrams.
We shall come back to this point below.
The third type of diagrams are 
the vertex corrections, 
which are by definition all diagrams contributing to the vertex function
$\Lambda^{\alpha} ( \tilde{k} ; \tilde{q} )$.
Naively, one could
try to expand the vertex function in powers of the 
RPA gauge field propagator,
 \begin{equation}
 \Lambda^{\alpha} ( \tilde{k} ; \tilde{q} )
 = 1 + \sum_{n=1}^{\infty} \Lambda_n^{\alpha} ( \tilde{k} ; \tilde{q} )
 \label{eq:verteexpansion}
 \; ,
 \end{equation}
where $\Lambda_n^{\alpha}$ is the sum of all contributions
to the three-legged vertex  involving  $n$ powers 
of $h^{{\rm RPA} , \alpha}$.
However, this expansion might be ill-defined, because successive powers
might be more and more singular. In this case non-perturbative methods are
necessary to resum the  perturbation series.
Here we shall restrict ourselves to the more modest task
of 
evaluating the leading vertex correction $\Lambda^{\alpha}_1$
shown in Fig.\ref{fig:vertex},  
which is explicitly given by
 \begin{equation}
 \Lambda_1^{\alpha} ( \tilde{k} ; \tilde{q} ) = - \frac{1}{\beta V} 
 \sum_{ \tilde{q}^{\prime}} h^{{\rm RPA}, \alpha}_{ \tilde{q}^{\prime}}
 G_0^{\alpha} ( \tilde{k} + \tilde{q}^{\prime} ) 
 G_0^{\alpha} ( \tilde{k} + \tilde{q} + 
 \tilde{q}^{\prime} )
 \; .
 \label{eq:lambda1}
 \end{equation}
An attempt to evaluate Eq.(\ref{eq:lambda1}) has been made 
in Ref.\cite{Altshuler94}. However, as will be explained 
in Sec.3.3, 
these authors have missed the dominant contribution.
Before embarking on a careful evaluation of Eq.(\ref{eq:lambda1}),
let us explain with the help 
of Ward-identities why vertex corrections can be
expected to play an important role in the present problem.

\subsection{Ward-identities}
\label{subsec:ward}
\noindent
In the limit of vanishing energy-momentum transfer 
$\tilde{q} = [ {\bf{q}} ,
i \omega ]$, the three-legged vertex
$\Lambda^{\alpha} ( {\tilde{k}} ; {\bf{q}} , i \omega )$ can be
related to appropriate partial
derivatives of the self-energy $\Sigma^{\alpha} ( \tilde{k} )$.
In the {\it{dynamic limit}} we first set ${\bf{q}} = 0$ and
then let $\omega \rightarrow 0$ (see Ref.\cite{Nozieres64}). 
The corresponding Ward-identity 
reads in our notation
 \begin{equation}
 \Lambda^{\alpha}_{\rm dy} ( {\tilde{k}} ) \equiv
 \lim_{\omega \rightarrow 0} \Lambda^{\alpha} ( {\tilde{k}} ; 0 , i \omega )
 = 1 - \frac{ \partial \Sigma^{\alpha} 
 ( \tilde{k} )}{\partial ( i \tilde{\omega}_n ) }
 \; .
 \label{eq:Warddy}
 \end{equation}
On the other hand, in the {\it{static}} limit, we set first $\omega = 0$ 
and then take the limit ${\bf{q}} \rightarrow 0$. This yields
  \begin{equation}
 \Lambda^{\alpha}_{\rm stat} ( {\tilde{k}} ) \equiv
 \lim_{{\bf{q}} \rightarrow 0} \Lambda^{\alpha} ( {\tilde{k}} ; {\bf{q}} , 0 )
 = 1 + \frac{ \partial \Sigma^{\alpha} 
 ( \tilde{k} )}{\partial ( v_F k_{\|} ) }
 \; .
 \label{eq:Wardstat}
 \end{equation}
It has been argued\cite{Stern95} that these Ward-identities 
imply that vertex corrections are  negligible in the present
problem. The argument is based on the observation that
the gauge field propagator (\ref{eq:hrpa})
is singular in the static limit
$| \omega | \ll v_F q$, so that at
the first sight it seems that the vertex correction should be estimated
from the static Ward-identity (\ref{eq:Wardstat}). Because the 
first order self-energy (\ref{eq:GWspherical6}) is
independent of $k_{\|}$, this would imply that the 
vertex $\Lambda^{\alpha} ( \tilde{k} ; \tilde{q} )$
in Eq.(\ref{eq:skeleton}) can be savely replaced by unity.
However, this argument is incorrect, because it ignores the fact that
the non-analyticity of the first order self-energy is entirely due
to momentum transfers that are essentially parallel to the Fermi surface.
To see this more clearly, let us recall that for
interactions with dominant forward scattering the following more general
Ward-identity can be derived\cite{Metzner97}, 
 \begin{equation}
 [ i \omega - {\bf{v}}^{\alpha} \cdot {\bf{q}} ]
 \Lambda^{\alpha} ( \tilde{k} ; \tilde{q} )
 = \frac{1}{G^{\alpha} ( \tilde{k} + \tilde{q} / 2 )}
 - \frac{1}{G^{\alpha} ( \tilde{k} - \tilde{q} / 2 )}
 \label{eq:Wardgen}
 \: .
 \end{equation}
The dynamic Ward-identity
Eq.(\ref{eq:Warddy}) can be obtained as
a special case of Eq.(\ref{eq:Wardgen}) by taking the limits
$\omega \rightarrow 0$ and ${\bf{q}} \rightarrow 0$ such that
the ratio
 \begin{equation}
 r \equiv \frac{  {\bf{v}}^{\alpha} \cdot {\bf{q}} }{ | \omega | }
  = \frac{  v_F q \cos \vartheta }{| \omega |} 
 \label{eq:rdef}
  \end{equation}
vanishes\cite{Metzner97}. Similarly, the static Ward-identity (\ref{eq:Wardstat})
is obtained by  taking these limits such that $r \rightarrow \infty$.
The crucial point is now that according to Eq.(\ref{eq:regimesmomtrans})
the ratio $ v_F q \cos \vartheta /| \omega |$ 
is {\it{small compared with unity}}
for the relevant energy-momentum transfers
that are responsible for the non-analyticities in the
first-order self-energy. Hence,  
the vertex function should be 
estimated from the {\it{dynamic}} Ward-identity (\ref{eq:Warddy}).
If we now substitute the
perturbative self-energy (\ref{eq:GWspherical6})
into the right-hand side of Eq.(\ref{eq:Warddy}), we see that
the vertex actually diverges as $ | \tilde{\omega}_n |^{
\frac{ 1 - \eta}{1 + \eta} }$ for $\tilde{\omega}_n \rightarrow 0$
(recall that $\eta > 1$).
It should be kept in mind, however, 
that the Ward-identity (\ref{eq:Wardgen})
has been  derived by linearizing the energy dispersion 
in the vicinity of the
Fermi surface, and therefore does contain information about 
{\it{curvature}} effects.
As already pointed out in Refs.\cite{Ioffe94,Altshuler94}
and discussed in detail below,
the curvature of the Fermi surface
plays a crucial role in the present problem, 
so that the Ward-identities (\ref{eq:Warddy}--\ref{eq:Wardgen})
{\it{do not have much predictive power.}} In particular,
for a spherical Fermi surface the above Ward-identities
{\it{cannot be used to obtain qualitative estimates
for the order of magnitude of vertex corrections}}.
We shall verify the correctness of this statement
in the next section by explicitly evaluating the
leading vertex correction.

\subsection{The leading vertex correction to the self-energy}
\label{subsec:lowest}
\begin{figure}
\epsfysize3.3cm 
\hspace{23mm}
\epsfbox{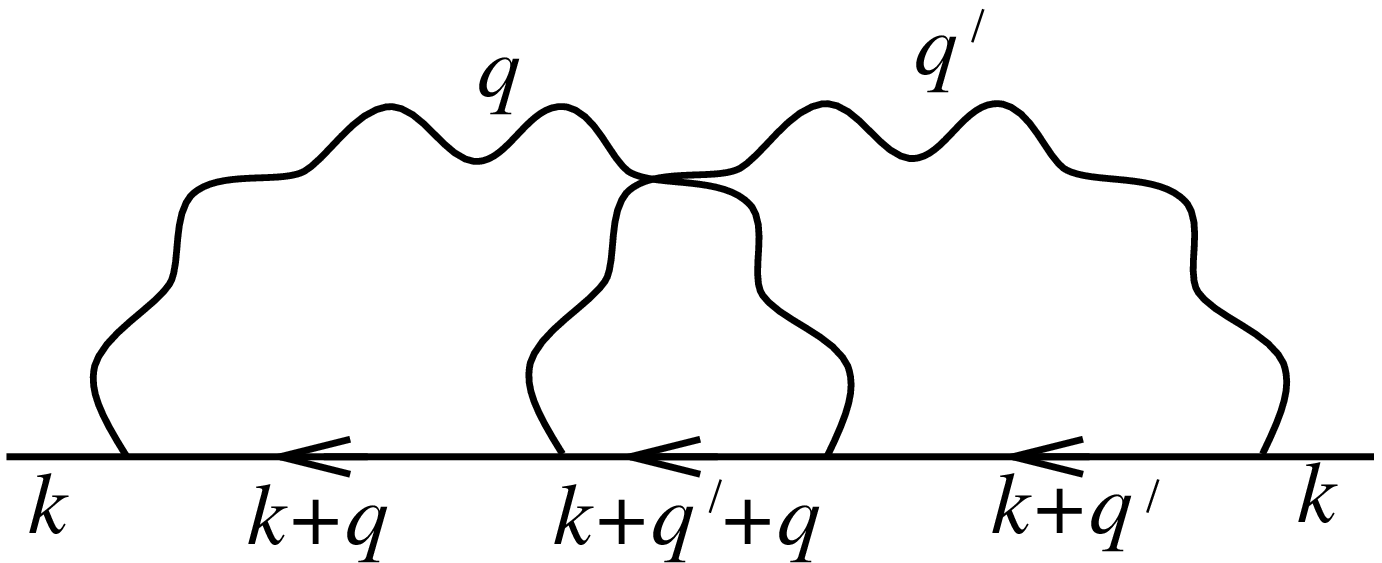}
\fcaption{Leading vertex correction to the
self-energy.}
\label{fig:Feynman2}
\end{figure}

\noindent
The vertex correction 
$\Lambda_1^{\alpha}$ in Eq.(\ref{eq:lambda1})
gives rise to the following
two-loop correction to the self-energy,
 \begin{equation}
 \Sigma^{\alpha}_2 ( \tilde{k} ) =
 - \frac{1}{\beta V} \sum_{ \tilde{q} }
 h^{ \alpha}_{\tilde{q}} 
 \Lambda^{\alpha}_1 ( \tilde{k} ; \tilde{q} )
 G^{\alpha}_0 ( \tilde{k} + \tilde{q} )
 \label{eq:sigma2}
 \;  .
 \end{equation}
The corresponding Feynman diagram is shown 
in Fig.\ref{fig:Feynman2}.
Note that in two dimensions the vertex $\Lambda^{\alpha}_1 ( \tilde{k} ; \tilde{q} )$
is a rather complicated function of six variables. 
From the evaluation of the
one-loop self-energy we expect that
the regime where the momentum transfers
${\bf{q}}$ and ${\bf{q}}^{\prime}$ are
almost parallel to the
Fermi surface will play an important role.
Below we shall first reproduce the result of 
AIM\cite{Altshuler94}, and then show
that these authors have missed the dominant contribution.
The reason is rather subtle: it turns out that
the dominant contribution to the {\it{self-energy}}
$\Sigma_2^{\alpha}$ is determined 
by a {\it{sub-dominant}} contribution to the {\it{vertex}}
$\Lambda_1^{\alpha}$.

\subsubsection{Exact manipulations}
\begin{figure}
\epsfysize4cm 
\hspace{23mm}
\epsfbox{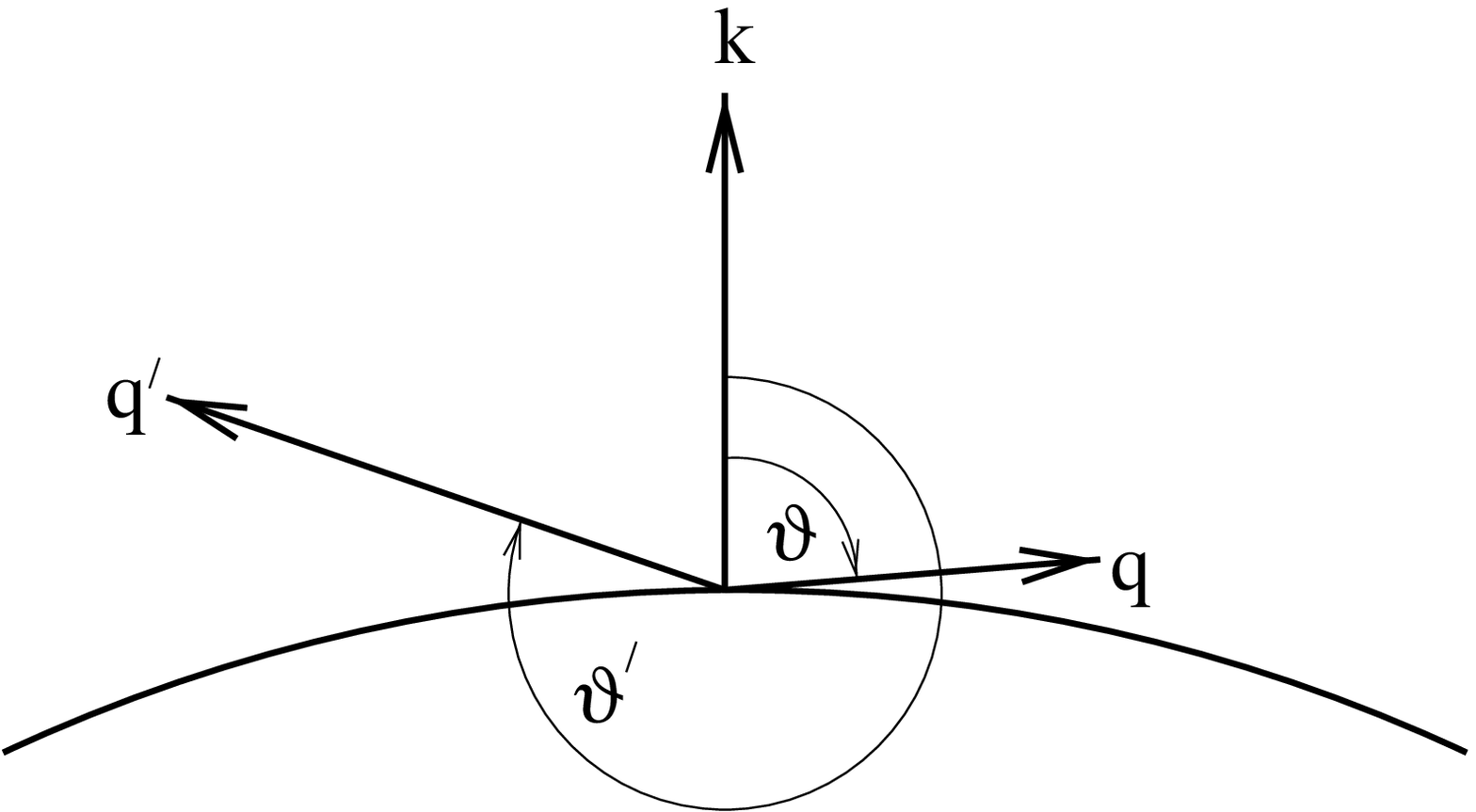}
\fcaption{Definition of the
angles $\vartheta$ and  $\vartheta^{\prime}$.}
\label{fig:coordinate2}
\end{figure}

\noindent
We begin with the calculation of the function
$\Lambda^{\alpha}_1 ( \tilde{k} ; \tilde{q} )$.
Setting for simplicity ${\bf{k}} = k_{\|} \hat{\bf{v}}^{\alpha}$ and
introducing the circular coordinates shown in Fig.\ref{fig:coordinate2}, 
we obtain
 \begin{eqnarray}
 \Lambda_1^{\alpha} ( k_{\|} , i \tilde{\omega}_n ; 
 q , \vartheta , i \omega)
 & = & 
 \nonumber
 \\  & & \hspace{-30mm}
\frac{1}{2 \pi v_F m^{\ast} ( 1 + \frac{k_{\|}}{k_c} )
 ( 1 + \frac{k_{\|} + q \cos \vartheta}{k_c} )}
 \int_{0}^{\kappa} d q^{\prime} 
 \int_{- v_F q^{\prime}}^{v_F q^{\prime}} d \omega^{\prime}
 \frac{ I ( W^{\prime} , U , \Delta) }{ \Gamma_{q^{\prime}} +
 | \omega^{\prime} |}
 \label{eq:Lambda1ex}
 \: ,
 \end{eqnarray}
with the dimensionless integral $I ( W^{\prime} , U , \Delta )$ given by
 \begin{equation}
 I ( W^{\prime} , U , \Delta ) = \frac{1}{2 \pi}
 \int_{- \pi}^{ \pi} d \vartheta^{\prime} 
 \frac{ \sin^2 \vartheta^{\prime} }{ ( W^{\prime} - \cos \vartheta^{\prime} )
 ( U - \cos \vartheta^{\prime} - \Delta \sin \vartheta^{\prime} )}
 \; ,
 \label{eq:Idef}
 \end{equation}
and
 \begin{equation}
 W^{\prime}  = 
  \frac{ G_0^{-1} + i \omega^{\prime} - \frac{ q^{\prime  2}}{2 
 m^{\alpha} }}{v_F q^{\prime} ( 1 + \frac{k_{\|}}{k_c} ) }
 \label{eq:Wprimedef}
 \; ,
 \end{equation}
  \begin{equation}
 U  = 
  \frac{ G_0^{-1} + i \omega^{\prime} + i \omega 
 - ( 1 + \frac{k_{\|}}{k_c} ) v_F q \cos \vartheta
 - \frac{ q^2 + q^{\prime  2}}{2 m^{\alpha} }  
 }{v_F q^{\prime} ( 1 + \frac{ k_{\|} + q \cos \vartheta }{k_c} ) }
 \label{eq:Udef}
 \; ,
 \end{equation}
 \begin{equation}
 \Delta = \frac{ q }{k_c} \frac{ \sin \vartheta}{ 
  ( 1 + \frac{ k_{\|} + q \cos \vartheta }{k_c} )}
 \; .
 \end{equation}
The angular integration in Eq.(\ref{eq:Idef}) can be performed exactly.
The result is
 \begin{eqnarray}
 I ( W^{\prime} , U , \Delta ) & = &
 \frac{4}{ (1 + W^{\prime})(1 + U) [ ( B - A)^2 + A C^2 ]}
 \nonumber
 \\
 &  & \hspace{-20mm} \times \left\{
 (B - A ) \left[ \frac{ 1 - \sqrt{A}}{ 1 - A} -
 \frac{ (1-B)( 1 - \sqrt{B - {C}^2 /4})}{
  ( 1 - B)^2 + {C}^2} 
 \right]
 \right.
 \nonumber
 \\
 &  & \hspace{-20mm} +
 \left.
 \frac{ {C}^2}{ ( 1 - B )^2 + {C}^2} \left[ - B +
 \frac{ ( B + A ) ( 1 + B)}{ 4 \sqrt{ B - {C}^2 / 4}}
  \right]
 \right\}
 \; ,
 \label{eq:Idone}
 \end{eqnarray}
where we have defined
 \begin{equation}
 A = \frac{ W^{\prime} - 1}{ W^{\prime} + 1}
 \; \;  ,  \; \; B = \frac{ U-1}{U+1} 
 \; \;  ,  \; \;
 C = \frac{ 2 \Delta}{U + 1}
 \; ,
 \end{equation}
and the roots have to be taken such that
 ${\rm Re} \sqrt{A} \geq 0$ and ${\rm Re} \sqrt{ B - C^2 / 4} 
 \geq 0$.
To make further progress without
resorting to numerical methods, the obvious strategy is to  identify the
regimes in the $q^{\prime}-\omega^{\prime}$-plane
which dominate the integral and then find some
simplification of the  integrand such that the integration
can be carried out analytically.
This is actually not so easy, because
we are eventually interested in the self-energy (\ref{eq:sigma2}), and
a priori we cannot exclude the possibility that,
due to certain symmetry-related cancellations,
sub-dominant contributions to the
vertex $\Lambda^{\alpha}_{1}$
are responsible for the dominant infrared behavior
of the self-energy $\Sigma_2^{\alpha}$. In fact, we shall
show shortly that this is precisely what happens.

\subsubsection{The contribution from the regime $q/ q^{\prime} = O (1)$}
\label{subsubsec:AIM}
\noindent
Let us for the moment assume that the 
leading infrared behavior of the self-energy
$\Sigma^{\alpha}_{2}$ in Eq.(\ref{eq:sigma2}) is determined by
the regime where
both parameters
$| W^{\prime} |  $ and $|U|  $ are small
compared with unity. At the first sight this
assumption seems reasonable, because in this case the integrand in
Eq.(\ref{eq:Idef}) has two poles in the vicinity of the real axis,
so that the value of the integral is large compared with unity.
We now show that with this assumption we can 
reproduce the  result of AIM\cite{Altshuler94}.
Later we shall demonstrate that this assumption is 
{\it{not correct}}: the infrared behavior of the self-energy in 
Eq.(\ref{eq:sigma2}) is in fact determined by the
regime where $|U| \gg 1$!

Obviously, the condition $| W^{\prime} | 
{ \raisebox{-0.5ex}{$\; \stackrel{<}{\sim} \;$}} 1$
is equivalent with (see also Eq.(\ref{eq:k0def}))
 \begin{equation}
 q^{\prime } { \raisebox{-0.5ex}{$\; \stackrel{>}{\sim} \;$}}  k_0
 = {\rm max} \left\{ | k_{\|} | , | \tilde{\omega}_n | / v_F \right\}
 \label{eq:Wprimesmall}
 \; ,
 \end{equation}
and $| U | { \raisebox{-0.5ex}{$\; \stackrel{<}{\sim} \;$}} 1$ requires
 \begin{equation}
 q^{\prime} { \raisebox{-0.5ex}{$\; \stackrel{>}{\sim} \;$}}  q_0
 = {\rm max} \left\{ k_0 ,  {| \omega | }/{ v_F} , 
 {q^2 }/{ k_c} \right\}
 \label{eq:Usmall}
 \; ,
 \end{equation}
where we have used the fact that according to Eq.(\ref{eq:regimesmomtrans})
$| \omega | { \raisebox{-0.5ex}{$\; \stackrel{>}{\sim} \;$}}
v_F q | \cos \vartheta |$.
In this regime it is easy to show from Eq.(\ref{eq:Idone}) that
to leading order
 \begin{eqnarray}
 I ( W^{\prime} , U , \Delta ) & \approx &
 \frac{i}{2} \left[ {\rm sgn } ( {\rm Im}  U )
 - {\rm sgn} ( {\rm  Im} W^{\prime} ) \right]
 \nonumber
 \\
 & \times &
 \left[ \frac{ 1}{ U - W^{\prime} - \Delta}
 + \frac{ 1}{ U - W^{\prime} + \Delta} \right]
 \label{eq:Iapprox1}
 \; .
 \end{eqnarray}
Actually, Eq.(\ref{eq:Iapprox1}) can be derived
in a much simpler way from Eq.(\ref{eq:Idef}) if we use the fact
that for small $| W^{\prime} |$ and $|U|$ the integral is dominated
by $\vartheta^{\prime} \approx \pm \pi / 2$. Then we may
substitute $\vartheta^{\prime} = x^{\prime} \pm \pi /2$, replace
$\sin (x^{\prime} \pm \pi / 2)  \approx\pm 1$, and
proceed as in Eq.(\ref{eq:intsmallx}). From
Eqs.(\ref{eq:Wprimedef}) and (\ref{eq:Udef}) 
we find to leading order 
 \begin{equation}
 U - W^{\prime} = \frac{ i \omega - v_F q \cos \vartheta - 
 \frac{ q^2}{2 m^{\alpha}} }{v_F q^{\prime}}
 \label{eq:UmiWprime}
 \; ,
 \end{equation}
and
 \begin{equation}
{\rm sgn } ( {\rm Im}  U )
 - {\rm sgn} ( {\rm  Im} W^{\prime} )   = 
 {\rm sgn} ( \tilde{\omega}_n + \omega + \omega^{\prime} )
 -
 {\rm sgn} ( \tilde{\omega}_n  + \omega^{\prime} )
 \; .
 \label{eq:Ims}
 \end{equation}
Substituting Eqs.(\ref{eq:Iapprox1}--\ref{eq:Ims}) into
Eq.(\ref{eq:Lambda1ex}), we obtain after a straightforward calculation
 \begin{eqnarray}
 \Lambda_1^{\alpha} ( k_{\|} , i \tilde{\omega}_n ; 
 q , \vartheta , i \omega)_{(q^{\prime} > q_0)}
 & = & 
 \nonumber
 \\
 & & \hspace{-50mm} 
\frac{-i }{2 \pi  m^{\ast} }
 \int_{q_0}^{\kappa} d q^{\prime}  q^{\prime}
 \left[ {\rm sgn} ( \tilde{\omega}_n )
 \ln \left( 1 + \frac{ | \tilde{\omega}_n |}{ \Gamma_{q^{\prime}} } \right)
 - {\rm sgn} ( \tilde{\omega}_n + \omega )
 \ln \left( 1 + \frac{ | \tilde{\omega}_n + \omega |}{ 
 \Gamma_{q^{\prime}} } \right) \right]
 \nonumber
 \\
 &  & \hspace{-50mm} \times
  \left[
 \frac{1}{ i \omega - v_F q \cos \vartheta - \frac{ q^2}{2 m^{\alpha}}
 - \frac{ q q^{\prime}}{m^{\alpha}} \sin \vartheta }
 +
\frac{1}{ i \omega - v_F q \cos \vartheta - \frac{ q^2}{2 m^{\alpha}}
 + \frac{ q q^{\prime}}{m^{\alpha}} \sin \vartheta }
 \right]
 \label{eq:Lambda1res}
 \: .
 \end{eqnarray}
The subscript $( q^{\prime} > q_0 )$
indicates that this is the contribution from
the regimes (\ref{eq:Wprimesmall}) and (\ref{eq:Usmall}). Note that
by definition $q_0 > k_0$.

Let us pause for a moment
and compare Eq.(\ref{eq:Lambda1res})
with the considerations of Sec.3.2.
Because the Ward-identity (\ref{eq:Wardgen}) ignores the
non-linear terms in the energy dispersion, let us
take the limit $1/ m^{\alpha} \rightarrow 0$ in Eq.(\ref{eq:Lambda1res}).
Comparing the $q^{\prime}$-integral with the corresponding
integral in the expression for $\Sigma_{1}^{\alpha}$
(see Eq.(\ref{eq:GWspherical4})),  and taking into
account that according to
Eq.(\ref{eq:regimesmomtrans}) 
$ | \omega |
{ \raisebox{-0.5ex}{$\; \stackrel{>}{\sim} \;$}}  
v_F q | \cos \vartheta | $ and 
that the integral is dominated
by frequencies 
$| \omega | { \raisebox{-0.5ex}{$\; \stackrel{<}{\sim} \;$}} 
| \tilde{\omega}_n |$,
it is easy so see that for linearized
energy dispersion 
$\Lambda_1^{\alpha}$ is indeed proportional to
$| \tilde{\omega}_n |^{ \frac{1 - \eta}{1 + \eta}}$. This
is in agreement
with the arguments presented in Sec.3.2:
in the limit $1/m^{\alpha} \rightarrow 0$ the order
of magnitude of the vertex
correction can be obtained  from the
{\it{dynamic}} Ward-identity (\ref{eq:Warddy}), and not
from the static Ward-identity (\ref{eq:Wardstat}).

However, the curvature terms in the denominator
of Eq.(\ref{eq:Lambda1res}) cannot be neglected!
Substituting Eq.(\ref{eq:Lambda1res})
into Eq.(\ref{eq:sigma2}), we obtain
 \begin{eqnarray}
 \Sigma_{2}^{\alpha} ( k_{\|} , i \tilde{\omega}_n )_{ ( q^{\prime} > q_{0})}
 & = & \frac{-i}{  (2 \pi )^3 ( m^{\ast })^2 v_F } \int_{0}^{\kappa} dq 
 \int_{ - v_F q}^{v_F q} d \omega \frac{1}{\Gamma_q + | \omega| }
 \nonumber
 \\
 & & \hspace{-42mm} \times 
 \int_{q_0}^{\kappa} d q^{\prime}  q^{\prime}
 \left[ {\rm sgn} ( \tilde{\omega}_n )
 \ln \left( 1 + \frac{ | \tilde{\omega}_n |}{ \Gamma_{q^{\prime}} } \right)
- {\rm sgn} ( \tilde{\omega}_n + \omega )
 \ln \left( 1 + \frac{ | \tilde{\omega}_n + \omega |}{ 
 \Gamma_{q^{\prime}} } \right) \right]
 \nonumber
 \\
 & & \hspace{-42mm} \times
 \int_{- \pi}^{\pi} d \vartheta \frac{\sin^2 \vartheta}{ W - \cos \vartheta }
 \left[ \frac{1}{ W_0 - \cos \vartheta - \frac{q^{\prime}}{k_c} \sin \vartheta}
 + \frac{1}{W_0 - \cos \vartheta + \frac{q^{\prime}}{k_c} \sin \vartheta} 
 \right]
 \; ,
 \label{eq:sigma2A}
 \end{eqnarray}
where $W$ is defined in Eq.(\ref{eq:Wdef}), and 
 $
 W_0 =  i \omega / ( v_F q ) - { q}/ (2  k_c )$.
The angular integration is the same as in Eq.(\ref{eq:Idef}), and can 
be done exactly. For $|W| \ll 1$ (which requires
$q > k_0$, see Eq.(\ref{eq:k0def})) the integral is dominated by
$\vartheta \approx \pm \pi / 2$, so that we may again set
$\sin \vartheta \approx \pm 1$ and extract the leading
behavior of the integral as in Eq.(\ref{eq:intsmallx}).
We obtain
  \begin{eqnarray}
 \Sigma_{2}^{\alpha} ( k_{\|} , i \tilde{\omega}_n )_{ ( q^{\prime} > q_{0})}
 & = &  \frac{- G_{0}^{-1} }{  2 \pi^2 ( m^{\ast })^2  } \int_{k_0}^{\kappa} 
 dq  q
 \int_{ 0}^{| \tilde{\omega}_n | } d \omega \frac{1}{\Gamma_q +  \omega }
 \nonumber
 \\
 &  & \hspace{-30mm} \times \int_{q_0}^{\kappa} d q^{\prime} q^{\prime} 
 \ln \left[ 1 - \frac{ \omega}{ \Gamma_{q^{\prime}} + | \tilde{\omega}_n | } 
 \right]
 \frac{ 1 }{ ( G_0^{-1} )^2 - 
 ( \frac{ q q^{\prime}}{m^{\alpha} } )^2 }
 \label{eq:sigma2Ares}
 \; .
 \end{eqnarray}
In the limit $1/ m^{\alpha} \rightarrow 0$ the term $q q^{\prime} / m^{\alpha}$
in the denominator of Eq.(\ref{eq:sigma2Ares}) vanishes, so that the integral
is proportional to $G_{0} = [ i \tilde{\omega}_n - \xi_{k_{\|}} ]^{-1}$. 
Then the dependence on $\tilde{\omega}_n$
can then be scaled out, and we find 
$\Sigma_{2}^{\alpha} / \Sigma_{1}^{\alpha}
\propto
 | \tilde{\omega}_n |^{ \frac{2}{1 + \eta}} / 
 ( i \tilde{\omega}_n - \xi_{k_{\|}})$. 
This rather singular result is drastically modified
by the curvature term $qq^{\prime} / m^{\alpha}$.
This is most easily seen by noting that according to Eq.(\ref{eq:qy}) the term
$qq^{\prime} / m^{\alpha}$ scales as 
$| \tilde{\omega}_n |^{\frac{2}{1 + \eta}}$,
which for $\eta > 1$  is much larger than $G_{0}^{-1}$.
To leading order we obtain
from Eq.(\ref{eq:sigma2Ares}) 
 \begin{equation}
 \Sigma_{2}^{\alpha} ( k_{\|} , i \tilde{\omega}_n )_{ ( q^{\prime} > q_{0})}
 \propto \left( \frac{ m^{\alpha}}{m^{\ast}} \right)^2 
 [ i \tilde{\omega}_n - \xi_{k_{\|}} ]
 \ln^2
 \left(  \frac{ q_c | \bar{\omega}_n |^{\frac{2}{1 + \eta}}  }{  k_{0} } \right)
 \; .
 \label{eq:sigma2Afinal}
 \end{equation}
For a circular Fermi surface, where $m^{\ast} = m^{\alpha}$,
this agrees with Eq.(8) of AIM\cite{Altshuler94}.
Obviously Eq.(\ref{eq:sigma2Afinal}) is negligible
compared with the first order self-energy
given in Eq.(\ref{eq:GWspherical6}). 

Because according to Eq.(\ref{eq:sigma2Afinal}) 
$| \Sigma_2 | \ll | \Sigma_1 |$, 
AIM argue that
the first order self-energy correction should actually be included
in all internal propagators.
In other words, the free Green's functions $G_{0}^{\alpha}$ in 
Eqs.(\ref{eq:lambda1}) and (\ref{eq:sigma2}) should be replaced by
the one-loop corrected Green' s function
$G_{1}^{\alpha}$.
If one repeats the above calculation with this renormalized
propagator, one finds\cite{Altshuler94} 
 $ \Sigma_{2}^{\alpha} \propto 
| \tilde{\omega}_n |^{ \frac{2}{1 + \eta} }$. This is much 
larger than Eq.(\ref{eq:sigma2Afinal}), 
and has the same order of magnitude as
$\Sigma_1^{\alpha}$.
However, we shall show shortly that the 
dominant infrared behavior of the two-loop
diagram
shown in Fig.\ref{fig:Feynman2} is 
not correctly given by Eq.(\ref{eq:sigma2Afinal}),
but is
in fact logarithmically larger
than the one-loop self-energy $\Sigma_1^{\alpha}$. 
Consequently, Eq.(\ref{eq:sigma2Afinal}) cannot
be used to justify the infinite resummation
of perturbation theory adopted 
in Ref.\cite{Altshuler94}, where all internal propagators
in higher order diagrams are replaced by one-loop corrected
propagators $G_1^{\alpha}$.
AIM  further justify
their approach by noting that the {\it{reducible}} diagram
corresponding to the term
$\Sigma_1 G_0 \Sigma_1$ in the first iteration
of the Dyson equation (see Fig.2b of Ref.\cite{Altshuler94})
is larger than the
{\it{irreducible}} diagram shown in Fig.\ref{fig:Feynman2}.
While such a point of view might be valid within 
the framework of a large-$N$ expansion,
we do not believe that for small $N$ the comparison
of irreducible and reducible diagrams is physically meaningful:
because reducible diagrams contain unphysical poles
for frequencies close to the non-interacting energy dispersion,
their contribution to 
the self-energy can always be tuned to be arbitrarily large.

\subsubsection{The contribution from the regime $q \gg q^{\prime}$}
\noindent
We now show that
in the limit of vanishing frequency 
the leading infrared behavior of
$\Sigma_2^{\alpha}$ defined in Eq.(\ref{eq:sigma2}) is not
given by Eq.(\ref{eq:sigma2Afinal}), 
but that the 
two-loop self-energy 
(\ref{eq:sigma2}) in fact exceeds 
the one-loop self-energy $\Sigma_{1}^{\alpha}$
by a logarithmically divergent factor,
 \begin{equation}
 \frac{ \Sigma_{2}^{\alpha} ( k_{\|} , i \tilde{\omega}_n )}{
 \Sigma_{1}^{\alpha} ( k_{\|} , i \tilde{\omega}_n ) }
 \sim - a_{\eta} \ln  ( 1/ | \bar{\omega}_n ) 
 \; \; \; \mbox{for $ |\bar{\omega}_n | \rightarrow 0$}
 \; ,
 \label{eq:sigma2final}
 \end{equation}
where $a_{\eta}$ is a positive numerical constant
proportional to $m^{\alpha} / m^{\ast}$.
The key observation is that so far
we have tacitly assumed
that the leading infrared behavior of Eq.(\ref{eq:sigma2})
is completely determined  by the regime 
where  $| U | { \raisebox{-0.5ex}{$\; \stackrel{<}{\sim} \;$}} 1$, 
which means that
$q^{\prime} { \raisebox{-0.5ex}{$\; \stackrel{>}{\sim} \;$}}  q_0
 = {\rm max} \left\{ k_0 ,  {| \omega | }/{ v_F} , 
 {q^2 }/{ k_c} \right\}$, see Eq.(\ref{eq:Usmall}).
Let us now check the contribution from the opposite
regime $| U | { \raisebox{-0.5ex}{$\; \stackrel{>}{\sim} \;$}} 1$, equivalent
with
$q^{\prime} { \raisebox{-0.5ex}{$\; \stackrel{<}{\sim} \;$}}  q_0$.
From Eq.(\ref{eq:Idone}) (or
more simply from Eq.(\ref{eq:Idef})) we see that in this case
 \begin{equation}
 I ( W^{\prime} , U , \Delta )  \approx 
 \frac{Z ( W^{\prime} )}{U} 
 \; \; \; , \; \; |U |
 { \raisebox{-0.5ex}{$\; \stackrel{>}{\sim} \;$}}  1
 \; ,
 \label{eq:IUlarge}
 \end{equation}
with $Z ( W^{\prime} )$ defined in Eq.(\ref{eq:Zdef}).
Note that $|U| { \raisebox{-0.5ex}{$\; \stackrel{>}{\sim} \;$}} 
 1$
implies $| \omega | { \raisebox{-0.5ex}{$\; \stackrel{>}{\sim} \;$}} 
 v_{F} q^{\prime}$, and hence 
$q \gg q^{\prime}$ (keeping in mind that
$v_F q
{ \raisebox{-0.5ex}{$\; \stackrel{>}{\sim} \;$}} | \omega |$).
Because by construction
 $|W^{\prime} | \ll 1$, it is clear
from Eqs.(\ref{eq:Udef}) and (\ref{eq:IUlarge}) 
that in this regime
 \begin{equation}
 I ( W^{\prime} , U , \Delta )  \approx 
 \frac{ - i {\rm sgn} ( \tilde{\omega}_n + \omega^{\prime} ) v_F q^{\prime} }{
 G_{0}^{-1} + i \omega - v_{F} q \cos \vartheta - \frac{q^2}{2 m^{\alpha}}
 }
 \; \; \; , \; \; q^{\prime} 
 { \raisebox{-0.5ex}{$\; \stackrel{<}{\sim} \;$}}  q_{0}
 \; .
 \end{equation}
Substituting this expression into Eq.(\ref{eq:Lambda1ex})
and performing the $\omega^{\prime}$-integration, we see that
the contribution from the regime
$q^{\prime} { \raisebox{-0.5ex}{$\; \stackrel{<}{\sim} \;$}}  q_{0}$
to the vertex $\Lambda_{1}^{\alpha}$ is
 \begin{eqnarray}
 \Lambda_1^{\alpha} ( k_{\|} , i \tilde{\omega}_n ; 
 q , \vartheta , i \omega)_{ (q^{\prime} < q_{0}) } & = &
  \frac{ - i}{\pi m^{\ast}} 
 \frac{ {\rm sgn} ( \tilde{\omega}_n ) }{
 G_{0}^{-1} + i \omega - v_F q \cos \vartheta - \frac{q^2}{2 m^{\alpha}} }
 \nonumber
 \\
 & \times &
 \int_{k_{0}}^{q_{0}} d q^{\prime} q^{\prime} 
 \ln \left( 1 + \frac{ | \tilde{\omega}_n | }{ \Gamma_{q^{\prime}}} 
 \right)
 \; .
 \label{eq:vertdom}
 \end{eqnarray}
Let us now assume that $| \omega | 
{ \raisebox{-0.5ex}{$\; \stackrel{>}{\sim} \;$}} 
 \Omega ( q , \vartheta ) $, where
 \begin{equation}
 \Omega ( q , \vartheta ) \equiv
 {\rm max} \{ v_{F} q \cos \vartheta , \frac{q^2}{2 m^{\alpha}} ,
 v_{F} q_c | \bar{\omega}_{n} |^{ \frac{1}{1 + \eta} } \}
 \; .
 \label{eq:omegacut}
 \end{equation}
The condition $|\omega | \geq
 v_{F} q_{c} | \bar{\omega}_{n} |^{ \frac{1}{1 + \eta} }$
allows us to neglect
the term $G_0^{-1}$ in the denominator of Eq.(\ref{eq:vertdom})
and to 
replace the upper cutoff of the $q^{\prime}$-integral
by infinity.
Introducing again the rescaled frequencies $\bar{\omega}_n
= \tilde{\omega}_n / ( v_F q_c )$ and
$\bar{\omega} = \omega / (v_F q_c )$ (see
Sec.2)
we obtain in this regime from Eq.(\ref{eq:vertdom})
 \begin{equation}
 \Lambda_1^{\alpha} ( k_{\|} , i \tilde{\omega}_n ; 
 q , \vartheta , i \omega)_{ (q^{\prime} < q_{0}) }
  \approx 
  - i c_{\eta} g \; {\rm sgn} ( \tilde{\omega}_{n} )
 \frac{  | \bar{\omega}_{n} |^{ \frac{2}{1 + \eta} } }{ i \bar{\omega} }
 \; \; \; , \; \; 
 | \omega | { \raisebox{-0.5ex}{$\; \stackrel{>}{\sim} \;$}} 
 \Omega ( q , \vartheta )
 \; ,
 \label{eq:vertodd}
 \end{equation}
where the numerical constant $c_{\eta}$ is defined
in Eq.(\ref{eq:FGWconst}).
Because by assumption $g \equiv q_c / k_F \ll 1$ and
$| \bar{\omega} | { \raisebox{-0.5ex}{$\; \stackrel{>}{\sim} \;$}} 
 | \bar{\omega}_{n} |^{ \frac{2}{1 + \eta} }$,
the numerical value of the 
vertex  function (\ref{eq:vertodd}) is certainly small
compared with  unity. Nevertheless, the contribution from this regime
dominates the infrared behavior of the self-energy, because
Eq.(\ref{eq:vertodd}) is an
{\it{odd function of $\omega$, and has therefore a different
symmetry than the bare vertex.}}
Substituting Eq.(\ref{eq:vertodd}) into
Eq.(\ref{eq:sigma2}) and using the fact that for $| \omega |
{ \raisebox{-0.5ex}{$\; \stackrel{>}{\sim} \;$}} \Omega ( q , \vartheta )$
we may approximate $G_{0}^{\alpha} ( \tilde{k} + \tilde{q} )
\approx ( i \omega )^{-1}$, we obtain
 \begin{eqnarray}
 \Sigma_{2}^{\alpha} ( k_{\|} , i \tilde{\omega}_n )_{ ( q^{\prime} < q_{0})}
 & \approx & \frac{v_F}{ ( 2 \pi )^2 m^{\ast} } 
 \int_{0}^{\kappa} dq q^2 \int_{- \pi}^{\pi} d \vartheta \sin^2 \vartheta
 \nonumber
 \\
 &  & \hspace{-32mm} \times 
 \int_{- v_F q}^{v_F q}
 d \omega
\frac{\Theta ( | \omega | - \Omega ( q , \vartheta  ))}{ i \omega
 ( \Gamma_{{q}} + | \omega | ) }
 \Lambda_1^{\alpha} ( k_{\|} , i \tilde{\omega}_n ; 
 q , \vartheta , i \omega)_{ (q^{\prime} < q_{0}) }
 \label{eq:sigma2dom}
 \; .
 \end{eqnarray}
Next we note that for sufficiently small $q$ the energy scale
$\Gamma_{q}$ is negligible compared with 
$q^2 / ( 2 m^{\alpha} )$, because $\Gamma_{q}$ vanishes faster
than $q^{2}$. Let us define the wave-vector $q_{1}$
where both energy scales are identical, 
 \begin{equation}
 \frac{ q_{1}^2}{ 2 m^{\alpha}} = \Gamma_{q_1} = v_{F} q_{c} 
 \left( \frac{q_{1}}{q_{c}} \right)^{ \frac{2}{1+ \eta} }
 \; ,
 \label{eq:q1def}
 \end{equation}
i.e.
  $ q_{1}  =  q_c 
 ( {q_{c}} / {2 k_c} )^{ 
 \frac{1}{\eta - 1}}$.
It turns out that  the leading contribution to
Eq.(\ref{eq:sigma2dom}) is  due to the 
regime $q < q_{1}$.
Because by construction $| \omega | > 
 \Omega ( q , \vartheta ) \geq q^2 / ( 2 m^{\alpha} )$, we may 
then neglect the energy $\Gamma_q$ compared with $| \omega |$ in
the integrand. The $\omega$-integration is now  trivial and 
generates a  factor of $\Omega^{-2} ( q , \vartheta )$.
Using  the expression (\ref{eq:GWspherical6})
for the one-loop
self-energy $\Sigma_{1}^{\alpha}$, we obtain
 \begin{equation}
 \Sigma_{2}^{\alpha} ( k_{\|} , i \tilde{\omega}_n )_{ ( q^{\prime} < q_{0})}
 = 
 \Sigma_{1}^{\alpha} ( k_{\|} , i \tilde{\omega}_n ) R ( i \tilde{\omega}_{n} )
 \; ,
 \label{eq:sigma21R}
 \end{equation}
with
 \begin{equation}
 R ( i \tilde{\omega}_{n} ) = - \frac{ v_F}{ ( 2 \pi )^2 m^{\ast}}
 \int_{0}^{q_{1}}
 dq q^2 \int_{- \pi}^{\pi} d \vartheta \frac{\sin^2 \vartheta}{ 
 \Omega^2 ( q , \vartheta) }
 \label{eq:Romegadef}
 \; .
 \end{equation}
To leading logarithmic order the integration in
Eq.(\ref{eq:Romegadef}) gives
 \begin{equation}
 R ( i \tilde{\omega}_{n} ) \sim - \frac{4}{\pi^2} \frac{ m^{\alpha}}{m^{\ast}}
 \ln ( q_{1} / q_{2} )
 \; ,
 \label{eq:R12}
 \end{equation}
where $q_{2}$ is defined by
 \begin{equation}
 \frac{q_{2}^2}{2 m^{\alpha}} = v_{F} q_{c} 
 | \bar{\omega}_{n} |^{\frac{2}{1 + \eta}} 
 \; .
 \end{equation}
Using Eq.(\ref{eq:q1def}),
we obtain for $| \bar{\omega}_n | \rightarrow 0$
  \begin{equation}
 R ( i \tilde{\omega}_{n} ) \sim  - \frac{a_{\eta}}{2} \ln ( 1/ |  \bar{\omega}_{n }  | )
 \; ,
 \label{eq:Rlog}
 \end{equation}
where  
 \begin{equation} 
 a_{\eta} = 
\frac{8}{\pi^2 ( 1 + \eta)} 
 \frac{ m^{\alpha}}{m^{\ast}}
 \; .
 \label{eq:aetadef}
 \end{equation}
Because of our rather crude method of estimating the integrals,
the numerical value of $a_{\eta}$ in Eq.(\ref{eq:aetadef}) should not
be taken too serious, and
remains uncertain by a factor
of the order of unity. 
Combining Eqs.(\ref{eq:sigma21R}), (\ref{eq:Rlog}) and
(\ref{eq:aetadef}), and taking into account that by symmetry
the regime $q^{\prime} \gg q$ gives rise to
an equally large contribution to the self-energy (this is
obvious from the labels 
in Fig.\ref{fig:Feynman2}), we finally arrive at
Eq.(\ref{eq:sigma2final}). 

Due to the complexity of the integrations, we have not been able
to check whether the result (\ref{eq:sigma2final}) is
modified if the internal free propagators $G^{\alpha}_0$ in Fig.\ref{fig:Feynman2}
are replaced by one-loop corrected propagators $G^{\alpha}_1$.
However, as discussed at the end of 
Sec.3.3.2, we do not believe that 
such an infinite resummation
of perturbation theory is a sensible procedure
(except perhaps within a large-$N$ expansion\cite{Altshuler94}).

\section{Conclusions}
\noindent
In this work we have shown that at the two-loop order the
low-frequency behavior of
the self-energy of fermions
that are coupled to transverse gauge fields with
propagator of the form (\ref{eq:hrpa}) is given by
 \begin{equation}
 {\Sigma ( i \tilde{\omega}_n )} = 
 - i c_{\eta} \frac{q^2_c}{m^{\ast}} {\rm sgn} ( \tilde{\omega}_n )
 \left| \frac{ \tilde{\omega}_n }{v_F q_c } \right|^{\frac{2}{1+ \eta}}
 \left[ 1 - a_{\eta} \ln \left( \frac{v_F q_c}{ | \tilde{\omega}_n | }
 \right) \right]
 \; ,
 \label{eq:selfresfinal}
 \end{equation}
where $c_\eta$ and $a_{\eta}$ are positive numerical constants.
The logarithmic correction in Eq.(\ref{eq:selfresfinal}) is 
due to
the leading radiative correction to the fermion - gauge
field vertex, and is
the main result
of this work. It implies that the
correct infrared behavior of 
the self-energy of fermions
that are coupled to transverse gauge fields
cannot be obtained from a one-loop
calculation. We disagree in this point with the authors
of Refs.\cite{Ioffe94,Altshuler94}, who  analyzed 
this problem
within a suitably devised $1/N$-expansion.

Because the leading vertex correction
generates an additional logarithm, it seems
likely that
the higher order vertex corrections
will give rise to even higher powers of logarithms.
If we boldly exponentiate the logarithmic
correction in Eq.(\ref{eq:selfresfinal}),
we find
 \begin{equation}
 {\Sigma ( i \tilde{\omega}_n )} = 
 - i c_{\eta} \frac{q^2_c}{m^{\ast}} {\rm sgn} ( \tilde{\omega}_n )
 \left| \frac{ \tilde{\omega}_n }{v_F q_c } \right|^{\gamma}
 \; , \; 
 \gamma = 
\frac{2}{1+ \eta} +
 a_{\eta}
 \; .
 \label{eq:selfresfinalexp}
 \end{equation}
Note that according to Eq.(\ref{eq:aetadef})
$a_{\eta}$ 
is positive and of the order of unity 
for a spherical Fermi surface.
It is therefore tempting to speculate that
the summation of vertex corrections to all orders will indeed
push the exponent $\gamma$ to a value that is
larger than predicted by the one-loop result. 
In fact, the non-perturbative calculation 
given in Ref.\cite{Kopietz97} suggests that the true 
exponent $\gamma$ is {\it{not}} smaller than unity, implying that 
the infrared fluctuations of the gauge field do
{\it{not}} lead to a destruction of the Fermi liquid.
This would explain
why experimentally
composite fermions in the half-filled Landau level
seem to behave as well-defined quasi-particles\cite{Willett93}.
%

\nonumsection{Acknowledgments}
\noindent
I have profited from a discussions with S. Simon
during a workshop on the quantum Hall effect at Villa Gualino, Torino,
which motivated me
to take a closer look at vertex corrections.
I would also like to thank
A. Maccarone for interesting discussions 
during his visit in G\"{o}ttingen,
and A. Millis and S. Chakravarty for
their comments on the manuscript.
This work was supported by a Heisenberg Fellowship of the
Deutsche Forschungsgemeinschaft.
 
\nonumsection{References}
\noindent
%
%


\begin{thebibliography}{99}
%
\bibitem[*]{address} 
Address from 1 November 1997 -- 15 April 1998.
%
\bibitem{Baskaran88}
G. Baskaran and P. W. Anderson, Phys. Rev. {\bf{B}} 37, 580 (1988).
%
\bibitem{Ioffe89}
L. B. Ioffe and A. I. Larkin, Phys. Rev. {\bf{B}} 39, 8988 (1989).
%
\bibitem{Lee89}
P. A. Lee, Phys. Rev. Lett. {\bf{63}}, 680 (1989);
N. Nagaosa and P. A. Lee, Phys. Rev. Lett. {\bf{64}},  2450 (1990).
%
\bibitem{Halperin93}
B. I. Halperin, P. A. Lee, and N. Read, Phys. Rev. {\bf{B}} 47, 7312 (1993).
%
\bibitem{Jain89}
J. Jain, Phys. Rev. Lett. {\bf{63}}, 199 (1989); Phys. Rev. 
{\bf{B}} 40, 
8079 (1989);
{{41}}, 7653 (1990).
%
\bibitem{Blok93}
B. Blok and H. Monien, Phys. Rev. {\bf{B}} 47, 3454 (1993).
%
\bibitem{Ioffe94}
L. B. Ioffe, D. Lidsky, and B. L. Altshuler,
Phys. Rev. Lett. {\bf{73}}, 472 (1994).
%
\bibitem{Altshuler94}
B. L. Altshuler, L. B. Ioffe, and A. J. Millis, Phys.
Rev. {\bf{B}} 50, 14048 (1994).
%
\bibitem{Khveshchenko93}
D. V. Khveshchenko and P. C. E. Stamp, Phys. Rev. Lett.
{\bf{71}}, 2118 (1993); Phys. Rev. {\bf{B}} 49, 5227 (1994).
%
\bibitem{Gan93}
J. Gan and E. Wong, Phys. Rev. Lett. {\bf{71}}, 4226 (1993).
%
\bibitem{Plochinski94}
J. Polchinski, Nucl. Phys. {\bf{B}} 422, 617 (1994).
%
\bibitem{Nayak94}
C. Nayak and F. Wilczek, Nucl. Phys. {\bf{B}} 430, 534 (1994).
%
\bibitem{Kwon94}
H. J. Kwon, A. Houghton, and J. B. Marston, Phys. Rev. Lett.
{\bf{73}}, 284 (1994);
Phys. Rev. {\bf{B}} 52, 8002 (1995).
%
\bibitem{Onoda95}
M. Onoda, I. Ichinose, and T. Matsui, 
Nucl. Phys. {\bf{B}} 446, 353 (1995);
Phys. Rev. {\bf{B}} 54, 13674 (1996).
%
\bibitem{Chakravarty95}
S. Chakravarty, R. E. Norton, and O. Syljuasen,
Phys. Rev. Lett. {\bf{75}}, 3584 (1995).
%
\bibitem{Kopietz96a}
P. Kopietz, Phys. Rev. {\bf{B}} 53, 12761 (1996).
%
\bibitem{Kopietz97}
P. Kopietz and G. E. Castilla, Phys. Rev. Lett. {\bf{78}}, 314 (1997).
%
\bibitem{Kopietz96}
P. Kopietz and G. E. Castilla, Phys. Rev. Lett. {\bf{76}}, 4777 (1996).
%
\bibitem{Kopietzbook}
P. Kopietz, {\it{Bosonization of Interacting Fermions 
in Arbitrary Dimensions}},
(Springer, Berlin, 1997).
%
\bibitem{Henrici88}
The complex function $Z ( W )$  defined in Eq.(\ref{eq:Zdef}) is
a particular inverse of the so-called {\it{Joukowski map}}
$Z \rightarrow W = \frac{1}{2} ( Z + \frac{1}{Z} )$, which is
uniquely defined by restricting the
$Z$-domain to the interior of the unit circle, see
P. Henrici, {\it{Applied and Computational
Complex Analysis, Vol.1}}, (Wiley Classics Library Edition, 1988),
pp. 294--298.
%
\bibitem{Stern95}
A. Stern and B. I. Halperin, Phys. Rev. {\bf{B}} 52, 5890 (1995).
%
\bibitem{Kopietz95}
P. Kopietz, J. Hermisson, and K. Sch\"{o}nhammer,
Phys. Rev. {\bf{B}} 52, 10877 (1995).
%
\bibitem{Metzner97}
W. Metzner, C. Castellani, and C. Di Castro,
preprint cond-mat/9702012,
Adv. Phys. (in press).
%
\bibitem{Kim94}
Y. B. Kim, A. Furusaki, X.-G. Wen, and P. A. Lee,
Phys. Rev. {\bf{B}} 50, 17917 (1994).
%
\bibitem{Nozieres64}
P. Nozi\`{e}res,
{\it{Theory of Interacting Fermi Systems}}, (Benjamin, New York, 1964).
%
\bibitem{Willett93}
R. L. Willett {\it{et al.}}, Phys. Rev. Lett. {\bf{71}}, 3846 (1993);
W. Kang {\it{et al.}}, {\it{ibid.}} 3850 (1993);
D. R. Leadley {\it{et al.}}, Phys. Rev. Lett. {\bf{72}}, 1906 (1994);
H. C. Manoharan {\it{et al.}}, Phys. Rev. Lett. {\bf{73}}, 3270 (1994);
R. R. Du {\it{et al.}}, Phys. Rev. Lett. {\bf{73}}, 3274 (1994);
P. T. Coleridge {\it{et al.}}, Phys. Rev. {\bf{B}} 52, R11603 (1995);
J. H. Smet {\it{et al.}}, Phys. Rev. Lett. {\bf{77}}, 2272 (1996);
R. L. Willett, Adv. Phys. {\bf{46}}, 447 (1997).
%


\end{thebibliography}
\end{document}